\documentclass[reprint,prc,onecolumn,preprintnumbers,superscriptaddress,longbibliography]{revtex4-2}
\usepackage{graphicx} 
\usepackage{orcidlink}
\usepackage{graphicx}
\usepackage{dcolumn}
\usepackage{bm}
\usepackage{textgreek}
\usepackage{placeins}
\usepackage{braket}
\usepackage{physics}
\usepackage{overpic}
\usepackage{amssymb}
\usepackage{amsmath}
\usepackage{subfigure}
\usepackage{qcircuit}
\usepackage{tikz}
\usepackage{adjustbox}
\usepackage{longtable}
\usepackage{soul}
\usepackage{tensor}
\usepackage{youngtab}
\usepackage{cleveref}

\usepackage[smalltableaux]{ytableau}

\usepackage{tikz}

\begin{document}

\title{Quantum Simulations of Two-Dimensional Non-Abelian Adjoint String Breaking}
\author{Anthony N. Ciavarella \,\orcidlink{0000-0003-3918-4110}}
\email{anciavarella@lbl.gov}
\affiliation{Physics Division, Lawrence Berkeley National Laboratory, Berkeley, California 94720, USA}
\affiliation{Applied Math and Computational Research Division, Lawrence Berkeley National Laboratory, Berkeley, California 94720, USA}

\author{Roland de Putter}
\email{Roland.de.Putter@ibm.com}
\affiliation{IBM Quantum, IBM Research - 1101 Kitchawan Rd, Yorktown Heights, NY, USA}

\author{Ed Younis \,\orcidlink{0000-0002-1306-1860}}
\email{edyounis@lbl.gov}
\affiliation{Applied Math and Computational Research Division, Lawrence Berkeley National Laboratory, Berkeley, California 94720, USA}

\author{Ermal Rrapaj \,\orcidlink{0000-0002-3222-7010}}
\email{ermalrrapaj@lbl.gov}
\affiliation{National Energy Research Scientific Computing Center, Lawrence Berkeley National Laboratory, Berkeley, California 94720, USA}

\date{\today}

\begin{abstract}
    Quantum computers offer the potential to directly probe the dynamics of strongly coupled quantum field theories. As a step towards reaching this potential, local Krylov-based truncations of a pure SU(2) lattice gauge theory on a triangular lattice are constructed. Adjoint strings connected to dynamical gluons are constructed in this truncated theory, and the resonances dominating the long-term dynamics at a large value of the gauge coupling are determined. Local operators are constructed to identify string oscillations and breakings. This is used to perform a quantum simulation of adjoint string breaking on an $8\times8$ and $16\times8$-site lattice with {\tt ibm\_boston} using all 156 qubits. Quantitative agreement with tensor network simulations is obtained for circuits with $7,634$ CZ gates with a two-qubit gate depth of $218$. In this simulation, the rates of oscillations and glueball production are identified with a distinctly non-Abelian signature of the underlying gauge group.

\end{abstract}

\maketitle

\section{Introduction}
At particle colliders, highly energetic pairs of quarks, anti-quarks, and gluons are produced in collisions. These particles are connected by strings of chromo-electric flux. As the pairs of particles separate, the amount of energy in the flux tube increases, which enables the pair production of more quarks and gluons. At long times, these quarks and gluons bind together to form the hadrons that are directly measured in detectors. This process of hadronization is non-perturbative and currently modelled by fitting semiclassical calculations to collider data~\cite{Sjostrand:2006za}. However, in principle, this process is described by quantum chromodynamics (QCD) and should be predictable from first principles.

Lattice quantum chromodynamics offers the ability to directly perform non-perturbative QCD calculations. Lattice QCD calculations on traditional compute resources have enabled the direct calculation of quantities such as hadron masses, form factors, and elastic scattering amplitudes. However, the simulation of generic real-time dynamics in lattice QCD suffers from an exponentially scaling sign problem on classical computers. Quantum computing offers the potential to avoid this sign problem and efficiently simulate real-time dynamics. This has motivated the development of encodings of Hamiltonian lattice gauge theories~\cite{Kogut:1974ag} onto discrete degrees of freedom that can be efficiently mapped onto quantum hardware. Most approaches have made use of an electric basis~\cite{Byrnes:2005qx,Banuls:2017ena,Klco:2018kyo,Banuls:2018jag,Zohar:2019ygc,Klco:2019evd,Paulson:2020zjd,Ciavarella:2021nmj,Kan:2021nyu,Kan:2021xfc,Zhang:2021bjq,Ciavarella:2021lel,Ciavarella:2022zhe,Davoudi:2022xmb,Zache:2023dko,Muller:2023nnk,Ciavarella:2023mfc,Farrell:2023fgd,Rigobello:2023ype,Sakamoto:2023cxs,Chai:2023qpq,Hariprakash:2023tla,Angelides:2023noe,Farrell:2024fit,Ciavarella:2024fzw,Rhodes:2024zbr,Magnifico:2024eiy,Guo:2024tnb,Gustafson:2024bww,Ciavarella:2024lsp,Cataldi:2025pja,Ciavarella:2025zqf,Ciavarella:2025bsg,Cataldi:2025rue,DiMarcantonio:2025cmf,Ciavarella:2025tdl,Balaji:2025yua,Chen:2026hnh}, which can be made more efficient through loop string hadron bases~\cite{Raychowdhury:2018tfj,Raychowdhury:2018osk,Raychowdhury:2019iki,Kadam:2022ipf,Kadam:2024ifg,Burbano:2024uvn,Kadam:2025trs,Ilcic:2025gel} or large $N_c$ expansions~\cite{Ciavarella:2024fzw,Ciavarella:2025bsg,Modi:2026syn}. Approaches using magnetic bases that use discrete subgroups~\cite{Hackett:2018cel,Lamm:2019bik,Alexandru:2019nsa,Alam:2021uuq,Ji:2022qvr,Gustafson:2022xdt,Gustafson:2023swx,Gustafson:2023kvd,Carena:2024dzu,Gustafson:2024kym,Assi:2024pdn,Kurkcuoglu:2024cfv} or employ gauge fixing~\cite{Bauer:2021gek,Grabowska:2022uos,Kane:2022ejm,DAndrea:2023qnr,Li:2024lrl,Grabowska:2024emw,Burbano:2024uvn,Froland:2025bqf,Froland:2026aff}, have also been developed. Alternative strategies use different Hamiltonians, such as quantum link models~\cite{Brower:1997ha,Brower:2003vy,Zache:2021ggw,Halimeh:2021ufh,Ciavarella:2022qdx,Osborne:2023rzx,Joshi:2025pgv,Cao:2026qky}, q deformed Hamiltonians~\cite{Zache:2023dko,Hayata:2023bgh,Hayata:2026xeo,Hayata:2026rmv,John:2026gut} and orbifolds~\cite{Buser:2020cvn,Bergner:2024qjl,Hanada:2025goy,Lamm:2026zzl}.

These theoretical developments have been used to perform the first quantum simulations of lattice quantum field theories. 
Large-scale simulations have been performed in one spatial dimension~\cite{Yang:2020yer,Su:2022glk,Farrell:2023fgd,Farrell:2024fit,Zemlevskiy:2024vxt,Ciavarella:2024lsp,Zhu:2024dvz,Liu:2024lut,Farrell:2025nkx,Schuhmacher:2025ehh,Davoudi:2025rdv,Chernyshev:2025lil,Chai:2025kbi,Xiang:2025qhq,Than:2025gso,Mark:2025wuo,Hudomal:2025gjv}, with limited results obtained in higher dimensions~\cite{Ciavarella:2021nmj,Ciavarella:2021lel,Mendicelli:2022ntz,Kane:2022ejm,Ciavarella:2023mfc,Kavaki:2024ijd,Gupta:2024gnw,Ciavarella:2024fzw,Gyawali:2024hrz,Crippa:2024hso,Kavaki:2025hcu,Cobos:2025krn,Saner:2025nrq,Hayata:2026rmv,Xu:2026ibi,Joshi:2026hfe,John:2026gut,Farrell:2026uac}. Many works on quantum simulations have studied string-breaking dynamics. In these simulations, the system is quenched with a $q\Bar{q}$ pair and the resulting dynamics are studied. These simulations have been performed in one and two spatial dimensions with both Abelian and non-Abelian gauge groups. $\mathbb{Z}_2$ lattice gauge theories have provided a testbed for probing several aspects of the string breaking process. These simulations have revealed a range of non-trivial real-time dynamics such as dynamical string breaking into particle–antiparticle pairs, often as a delayed, two-stage process and enhanced near resonance \cite{Verdel:2019chj, De:2024smi, Cochran:2024rwe, Xu:2026ibi}, the formation of confined mesonic bound states and glueball excitations \cite{Liu:2018fza, Xu:2026ibi}, confinement-induced breakdown of thermalization with suppressed entanglement growth and long-lived (weakly thermalizing) coherent oscillations \cite{James:2018qly, Robinson:2018wbx, Lin:2016egw, Lerose:2019jrs}, and even ballistic plasma formation with long-time memory at high energies \cite{Mark:2025wuo}.
Tensor network based simulations of string breaking dynamics have been performed for the Schwinger model~\cite{Liu:2024lut, Grieninger:2026bdq, Crippa:2024hso, Artiaco:2025qqq, Cao:2026qky, Joshi:2026hfe, Surace:2019dtp, Xiang:2025qhq, Zhou:2021kdl}. These simulations demonstrate that thermalization and the dynamics of entanglement play a non-trivial role in the dynamics of hadronization. Preliminary studies have been performed in non-Abelian lattice gauge theories~\cite{Ciavarella:2024lsp, Gupta:2026tcg, Cataldi:2025cyo, John:2026gut}.

Most efforts in applying quantum simulation to hadronization physics have focused on initial states containing a $q\Bar{q}$ pair. However, at hadron colliders, it is also possible to create pairs of gluons. In traditional hadronization models, gluons are modeled as effective color–anticolor pairs, so that hadronization proceeds via fundamental string fragmentation. As a result, genuine adjoint flux tubes and their breaking are not explicitly modeled.  Quantum simulations of adjoint strings would enable a direct assessment of the assumptions underlying these modeling approaches. It is expected that adjoint strings exhibit qualitatively distinct dynamics compared to $q\Bar{q}$ strings, most notably the possibility of gluon-mediated screening, the formation of gluelumps, and the absence of asymptotic string stability. These effects could manifest in observables sensitive to color flow and hadronization, including jet substructure variables (such as pull angles and interjet radiation), differences between quark and gluon jet multiplicities, and event-wide measures such as charged particle multiplicity and underlying event activity.  Preliminary studies have been performed simulating the dynamics of an adjoint string with static endpoints on a plaquette chain~\cite{John:2026gut}.

Performing quantum simulations at scale to inform these experiments will require both algorithmic and theoretical developments. In this work, a Krylov-based electric truncation is introduced for a $2+1D$ SU(2) gauge theory (without fermions) on a triangular lattice. This results in an efficient encoding of the truncated theory onto qubits. Simulations of the dynamics of an adjoint string with dynamical endpoints are performed using both tensor networks and quantum computers. The encoding onto the quantum computer is achieved in two ways: a geometric approach with manually constructed circuits and a compiler approach that synthesizes the time evolution operator on tiles of the lattice.

\section{SU(2) Lattice Gauge Theory}

\subsection{Truncated theory}

Most work on quantum simulation of lattice gauge theories makes use of the Kogut-Susskind Hamiltonian defined on a square lattice~\cite{Kogut:1974ag}. The Hamiltonian is given by
\begin{equation}
    \hat{H} = \frac{g^2}{2} \sum_l \hat{E}^2_l - \frac{1}{2g^2} \sum_p \left(\hat{\Box}_p + \hat{\Box}^\dagger_p \right) \, ,
\end{equation}
where $g$ is the gauge coupling, $\hat{E}^2_l$ is the electric energy on link $l$, and $\hat{\Box}_p$ is the trace of the product of parallel transporters around plaquette $p$. In this work, we will work in lattice units where the lattice spacing $a=1$. For a square lattice, the plaquettes correspond to squares whose corners are all nearest neighbors. In the context of quantum simulation, it has been found that using alternative lattice geometries such as a hexagonal or triamond lattice can offer computational advantages~\cite{Raychowdhury:2018tfj,Muller:2023nnk,Kavaki:2024ijd,Turro:2024pxu,Kavaki:2025hcu,Illa:2025dou,Illa:2025njz,Yao:2025cxs,Chen:2026hnh}. In this work, a triangular lattice will be used in two spatial dimensions to enable an efficient mapping to IBM's Heron quantum processors. The Kogut-Susskind Hamiltonian on a triangular lattice is identical to the square lattice, except now the plaquette operators are given by products of parallel transporters going around triangles instead of squares.

\begin{figure}
    \centering
    \includegraphics[width=\linewidth]{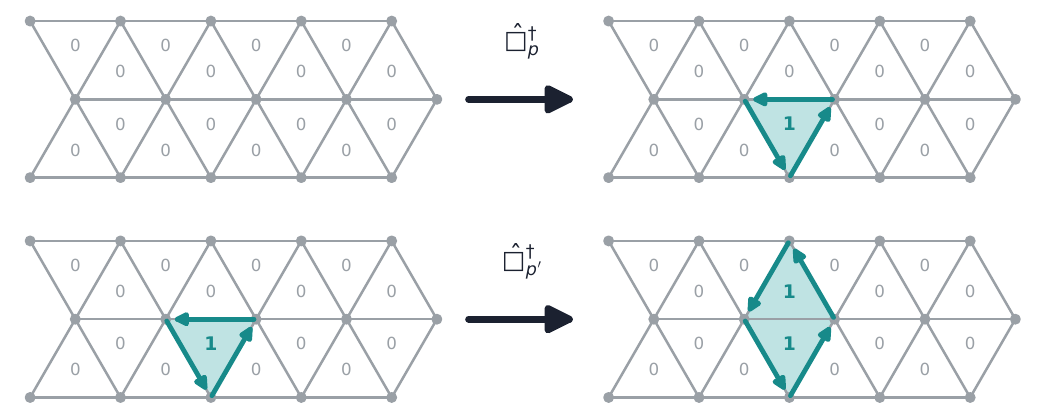}
    \caption{Mapping of the SU($2$) link gauge fields onto qubits placed at the center of each plaquette. The application of a plaquette operator flips a qubit from the $0$ to $1$ state. When neighboring qubits are in the $1$ state, their shared link is unexcited.}
    \label{fig:TruncatedLattice}
\end{figure}

The gauge degrees of freedom in the Kogut-Susskind Hamiltonian are infinite-dimensional and require a truncation to be mapped onto the finite memory of a quantum computer. Truncations in the electric basis result in a local Hamiltonian that maintains gauge invariance and have a factorial convergence to the untruncated theory~\cite{Ciavarella:2025tdl}. Recent work has extended electric basis truncations to include a local Krylov basis truncation where basis states are generated by repeated applications of operators from the Hamiltonian to a reference state~\cite{Ciavarella:2025bsg,Modi:2026syn}. With the harshest truncation, electric basis states can be represented using a single qubit per plaquette. Regions of neighboring qubits in the $\ket{1}$ state correspond to links on the boundary of that region having electric flux in the $j=1/2$ irrep. If all qubits neighboring a link are in the $\ket{0}$ state, then the link carries the $j=0$ irrep. Example basis states and transitions generated by the plaquette operator are shown in Fig.~\ref{fig:TruncatedLattice}. The Hamiltonian at this truncation is given by 
\begin{align}
    \hat{H} &= \hat{H}_E + \hat{H}_B \nonumber \\
    \hat{H}_E &= g^2 \frac{3}{8} \sum_p \left( 3\hat{P}^1_p - \sum_{\hat{n}} \hat{P}^1_p \hat{P}^1_{p+\hat{n}}\right) \nonumber \\
    \hat{H}_B &= -\frac{1}{g^2} \sum_p \left(\prod_{\hat{n}} \hat{C}_{p+\hat{n}}\right) \hat{X}_p \nonumber \\
    \hat{P}^q_p &= \ket{q}_p \bra{q}_p \nonumber \\
    \hat{C}_p &= \hat{P}^{0}_p + \frac{1}{2} \hat{P}^{1}_p \ \ \ ,
    \label{eq:SU2Ham}
\end{align}
where $p$ sums over all plaquettes on the lattice and $p+\hat{n}$ corresponds to the neighboring plaquette in the $\hat{n}$ direction. See Appendix~\ref{sec:Hamiltonian} for details. The electric energy on a link (neighboring plaquettes $p_1$ and $p_2$) is given by 
\begin{equation}
    \hat{E}^2 = \frac{3}{4} \left(\hat{P}^1_{p_1} \hat{P}^0_{p_2} + \hat{P}^0_{p_1} \hat{P}^1_{p_2}\right) = \frac{3}{8}\left(1 - \hat{Z}_{p_1} \hat{Z}_{p_2}\right) \ \ \ .
\end{equation}

\subsection{String phenomenology}
\label{subsec:string_pheno}

Studying string breaking in this truncated theory requires the creation of a string of electric flux. Previous simulations of string breaking on quantum computers to date have focused on the case where a quark-antiquark pair is connected by a string of electric flux. The initial $q\Bar{q}$ pair is created on the lattice by applying an operator of the form 
\begin{equation}
    S_{q\Bar{q}}(x,y) = \psi^\dagger(x) \left(\prod_{l\in \mathcal{L}} U_l\right) \psi(y) \, ,   
\end{equation}
where $\psi^{\dagger}(x)$ creates a quark at position $x$, $\psi(y)$ creates an anti-quark at position $y$, and $\mathcal{L}$ is a line of links that starts at $x$ and ends at $y$. However, in high-energy collisions, it is possible to produce a pair of gluons that undergo a similar string-breaking process. In the continuum, a gluon can be created by applying $A_{\mu}(x)$ to the vacuum. On the lattice, $A_{\mu}(x)$ is related to the link operator at position $x$ pointing in direction $\mu$ by $U_{\mu}(x) = e^{i A_{\mu}(x)}$. Therefore, one can create a string ending in a pair of gluons by applying the operator
\begin{equation}
    S_{gg} = U_{\mu}(x) \left( \prod_{l \in \mathcal{L}_1} U_l\right) \left( \prod_{l \in \mathcal{L}_2} U_l\right) U_{\nu}(y) \, ,
\end{equation}
where $\mathcal{L}_1$ is a line connecting $x$ to $y$ and $\mathcal{L}_2$ is a line connecting $x+\mu$ to $y+\nu$. In this truncated theory, applying this operator will create a loop of electric flux on the lattice. Therefore, closed loops of electric flux can be interpreted as a string that terminates in a dynamical pair of gluons. Denoting the set of qubits contained inside the loop by $\mathcal{A}$, and the set of qubits directly neighboring the exterior of the loop by $\partial \mathcal{A}$, $S_{gg}$ is given by
\begin{equation}
    \label{eq:string_creation_op}
    S_{gg} = \left(\prod_{q\in \partial \mathcal{A}} \hat{C}_q\right)\left(\prod_{q\in \mathcal{A}} \hat{X}_q\right) \ \  \ ,
\end{equation}
in this truncation of the theory.

In previous quantum simulations of string breaking performed for lattice gauge theories with matter, it was found that the dominant string breaking behavior was determined by resonance conditions set by the relation between the gauge coupling $g$, and the fermion mass $m$~\cite{Liu:2024lut, Cochran:2024rwe, Xu:2025abo}. On a coarse lattice (large $g$), the behavior of the adjoint string is also determined by resonances; however, there is no fermion mass to vary in this theory. While this limit is far from continuum physics, it can be used to develop qualitative insight into the dynamics of the theory. In the large $g$ limit, the simplest adjoint string states correspond to a line of plaquettes with a loop of electric flux in the fundamental representation flowing around it, and no other electric fields excited on the lattice. An example of an adjoint string is shown in the left panel of Figure~\ref{fig:string_configs}. As discussed in Appendix~\ref{sec:rates}, the evolution of electric basis states in this regime can be approximated using time-dependent perturbation theory, where the free part of the Hamiltonian is taken to be the electric Hamiltonian. The electric energy of a state is directly proportional to the number of excited links. The dynamics will be dominated by resonant transitions that keep the number of excited links in the lattice conserved. 

The lowest-lying excitation has three links excited, forming a triangle on the lattice. This state corresponds to a single glueball. The dynamics of a state with a glueball will be determined by second-order resonances that allow the glueball to move throughout the lattice. Longer loops of electric flux correspond to adjoint strings. Strings with a number of links divisible by $3$ are on resonance with states where there are only loops flowing around at most one plaquette each, and at long times will evolve to field configurations of this form. This could physically be interpreted as a gas of glueballs. Note that adjoint strings that enclose only two or three neighboring plaquettes have only $4$ or $5$ links excited, and so they cannot decay to glueballs. These states can be interpreted as being stable excited states of glueballs. The top left field configuration of Figure~\ref{fig:string_configs} shows an adjoint string where a glueball has broken off of the adjoint string, and the top right shows an excited glueball breaking off of the string. In addition to states where the string breaks and glueballs are emitted, an adjoint string can be on resonance with other states with a single loop of electric flux of the same length. Evolving to these states would correspond to the string oscillating between different configurations. The bottom row of Figure~\ref{fig:string_configs} shows different string oscillations that are on resonance with the initial state on the left. As shown in Appendix~\ref{sec:rates}, at leading order in time-dependent perturbation theory, the on-resonance matrix element for string oscillations cancels, while the matrix element for string-breaking transitions does not. Therefore, the long-time dynamics will be dominated by string-breaking behavior, with oscillations suppressed. However, at short times, the matrix element for string oscillations is twice the matrix element for string breaking. This indicates that there must be a cross-over between oscillation-dominated dynamics and string-breaking-dominated dynamics. This crossover is a direct consequence of the non-Abelian nature of the SU(2) gauge group. This hierarchy of timescales—a short-time regime dominated by coherent oscillations followed by a long-time regime dominated by breaking into glueballs—has no analog in Abelian gauge theories, where the color structure responsible for the separation in timescales is absent.
\begin{figure}
    \centering
    \includegraphics[width=0.5\linewidth]{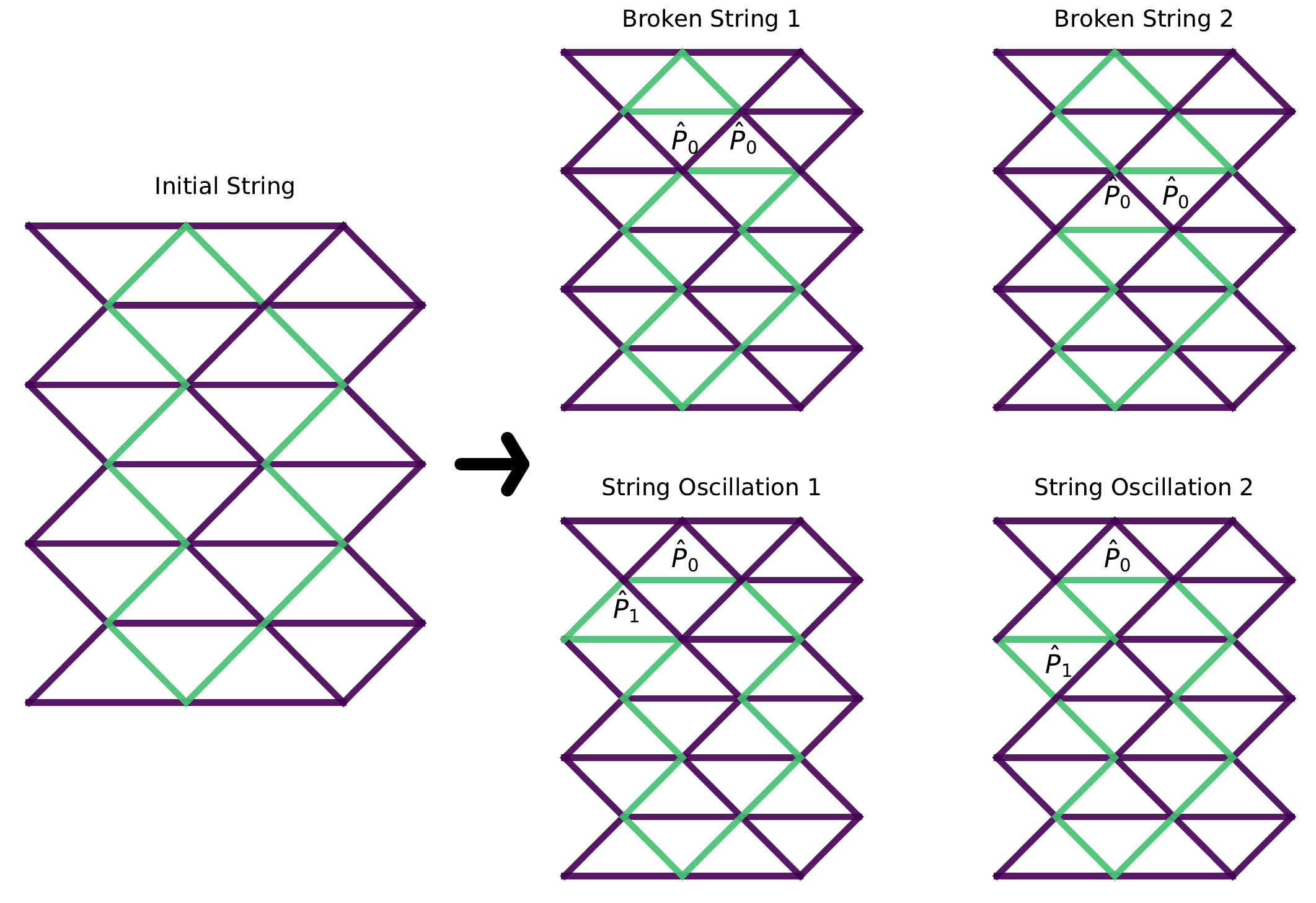}
    \caption{Different possible string configurations on the lattice. The left panel shows a single adjoint string. The top row on the right shows different ways this adjoint string can break on resonance, and the bottom row shows different ways that the string can oscillate. Note that there are other allowed final states given by rotations and reflections of these transitions. The probability for each configuration can be determined by measuring the pair of projectors shown on the lattice.}
    \label{fig:string_configs}
\end{figure}

Distinguishing between the possible dynamics of the string in a quantum simulation requires the construction of appropriate observables. One possible option is to measure the probability of different string configurations in the lattice. Explicitly, for a string $S$, the operator that projects onto the state where $S$ is present in the system is given by
\begin{equation}
    \hat{\Pi}_S = \left(\prod_{q\in S} \hat{P}^{1}_p \right) \left(\prod_{q\in \partial S} \hat{P}^{0}_p \right) \ \ \ ,
\end{equation}
where $S$ denotes the set of qubits within the area enclosed by the string and $\partial S$ is the set of qubits directly neighboring the outside of the string. While the theoretical interpretation of $\hat{\Pi}_S$ is clear, it is a high-weight operator that can be highly sensitive to errors that occur on a quantum computer. 

Alternatively, one can use lower-weight operators to distinguish different types of dynamics during the early evolution of the string. For example if $r$ and $r+\hat{x}$ are plaquettes inside the initial string, the operator $\hat{B}_r \equiv \hat{P}^{0}_{r}\hat{P}^{0}_{r+\hat{x}}$ will begin with expectation value $0$ and takes the value $1$ for states where the string breaks between $r$ and $r+\hat{x}$. Measuring this operator tells us how likely the string is to have broken at a given time and position. For the string to oscillate without breaking, a plaquette at the end of the string has to be de-excited while the plaquette next to the end stays excited. This transition can be tracked by measuring $\hat{O}_r \equiv\hat{P}^{0}_{r}\hat{P}^{1}_{r'}$ at each end of the string, where $r$ is the plaquette at the end of the string and $r'$ is the plaquette being excited by the string oscillation. In this work, we will use the expectation values $\langle \hat{B}_r \rangle$ and $\langle \hat{O}_r \rangle$ to quantify the string breaking and oscillation probabilities.

\section{Simulation methodology}
\label{sec:sim_method}

We simulate string evolution for the Hamiltonian in Eq.~(\ref{eq:SU2Ham}), where each qubit describes a plaquette on a triangular lattice, in both the strong ($g=1.4$) and intermediate ($g=1.1$) coupling regimes. Our high-level workflow is to prepare an initial string state, evolve it with the Hamiltonian using a Trotter approximation, and then measure electric link energies, string-breaking probabilities, and string-oscillation probabilities.
All quantum simulations are performed on {\tt ibm\_boston}, an IBM Heron r3 device with 156 qubits connected according to a heavy-hex topology.

We consider two methods for mapping the triangular lattice of plaquettes to the heavy-hex lattice. The first method is a ``geometric encoding'' (see Appendix~\ref{sec:ManualTrotter} and Fig.~\ref{fig:HeronMapping}), where plaquettes are mapped to the vertices of connectivity 3 on the heavy-hex lattice (i.e. the vertices of the hexagonal lattice embedded in the heavy-hex lattice) and interactions between plaquettes are mediated by ancilla qubits connecting the plaquettes (i.e.~the additional qubits on each edge of the hexagonal lattice that turn the hexagonal into a {\rm heavy}-hexagonal coupling map). Second-order Trotter evolution is implemented by alternating between evolution on upward (blue in Fig.~\ref{fig:HeronMapping}) and downward (green) pointing triangles (see Appendix~\ref{sec:ManualTrotter} for details). This approach allows allows us to simulate an $8 \times 8$ lattice of plaquettes, using a total of 148 qubits. While this mapping has a large qubit overhead, it enables the construction of relatively low-depth circuits to perform time evolution. Another advantage is that the ancilla qubits can be used to detect errors and remove faulty shots: after mediating a step of plaquette evolution, the ancillas should return to the zero-state; for a given plaquette operator, we therefore discard samples where any of the neighboring ancillas qubits are measured in the one-state.

The second mapping is a ``dense encoding'' (Appendix~\ref{sec:tiledtrotter}).
The geometric encoding was based on mapping the lattice geometry to the geometry of the quantum processor. Quantum compilers can enable the construction of less intuitive circuits; however, they are only able to act directly on a limited number of qubits. Appendix~\ref{sec:tiledtrotter} describes how one can compactly map the triangular lattice onto {\tt ibm\_boston} and tile the lattice with blocks that a quantum compiler can act on individually. For this encoding BQSKit~\cite{younis2021berkeley} was employed to synthesize a first-order Trotterized time evolution operator for a $16 \times 8$ lattice using all 156 qubits on {\tt ibm\_boston}. This compilation has a space-time tradeoff with larger circuit depths required per Trotter step. We have verified that the Trotter error is minimal in both encodings for the chosen simulation parameters.

The initial string state is prepared by first initializing (an approximation to) the vacuum state and then applying a version of Eq.~(\ref{eq:string_creation_op}). The details differ between the strong- and intermediate coupling simulations and are be explained in the respective sections. We create a vertical string containing 10 plaquettes, which we map to the right-hand side of the lattice, but away from the borders, in order to avoid a qubit that consistently has below-average performance (qubit 85).
The output of our quantum simulations are expectation values of the electric link energies $\hat{E}_l$, string breaking probabilities $\hat{B}_r$, and string oscillation probabilities $\hat{O}_r$. All of these operators are estimated from measurements in the computational basis (i.e.~no basis changes are needed). Before executing on {\tt ibm\_boston}, circuits are transpiled to the native gates of the backend, namely CZ, X, SX and RZ.

Errors are suppressed and mitigated using a combination of dynamical decoupling~\cite{Viola:1998gg}, Pauli twirling~\cite{Wallman:2015uzh}, twirled readout error extinction (TREX)~\cite{Berg:2020ibi}, operator decoherence renormalization (ODR)~\cite{Urbanek:2021oej,ARahman:2022tkr,Farrell:2023fgd}, zero noise extrapolation (ZNE)~\cite{Urbanek:2021oej}, averaging using reflection symmetry, and post-selection on ancilla qubits (see Appendix \ref{sec:Mitigation} for details).
For comparison with the quantum simulations, classical simulations are performed using matrix product state (MPS) tensor networks with bond dimensions up to $540$ to guarantee convergence.  We refer to Appendix \ref{sec:TensorSim} for a detailed study of tensor network simulation convergence, and a comparison between MPS and projected entangled pair states (PEPS) using belief propagation.

\section{Strong Coupling Regime}
\label{sec:strongsim}

\subsection{Geometric Encoding}
\label{subsec:geo_encoding}

\begin{figure}
    \centering
    \includegraphics[width=\linewidth]{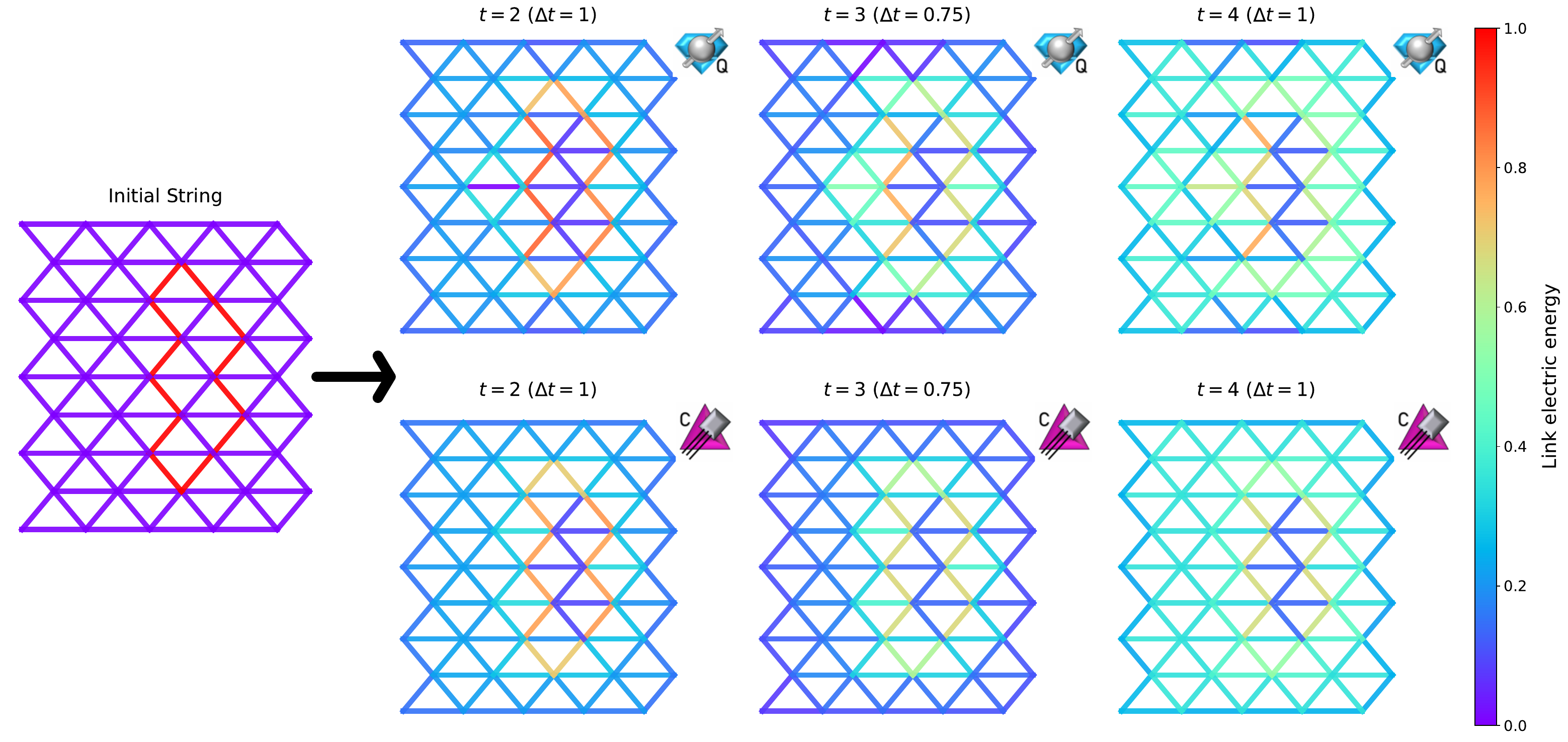}
    \caption{Evolution of an adjoint string on an $8\times8$ lattice with $g=1.4$. The upper plots show the error-mitigated results obtained from {\tt ibm\_boston} and the lower plots show the results of an MPS simulation with maximum bond dimension $100$. This evolution shows the breaking of the string and energy beginning to flow away from the initial string configuration.}
    \label{fig:string_ev_8x8_g14}
\end{figure}

As a probe of the strong coupling dynamics, simulations were performed with $g=1.4$ on an $8\times8$ lattice, mapping the triangular lattice to the hardware's heavy-hexagonal lattice using the geometric encoding.
As estimated in Appendix~\ref{sec:truncation}, the leading-order probability of a link leaking out of the truncated Hilbert space at this coupling is bounded by $0.095$, indicating that the truncated theory should provide a quantitatively reliable approximation to the untruncated dynamics over the timescales simulated here. The system was initialized with all qubits in the zero state. As discussed in Appendix~\ref{sec:VacPrep}, the all-zero state is close to the true vacuum for this coupling. An adjoint string was placed on the lattice by flipping a line of qubits into the one state, as shown in Fig.~\ref{fig:string_ev_8x8_g14}.


Fig.~\ref{fig:string_ev_8x8_g14} shows the electric energy on each link in both the MPS simulation and the quantum simulation up to a total evolution time of $t=4$\footnote{This figure shows the link energy with the normalization $\hat{L}=\frac{1}{2}\left(1 - \hat{Z}_{p_1} \hat{Z}_{p_2}\right)$, so that it is bounded by $[0, 1]$, which differs from the previously defined $\hat{E}^2$ by a constant factor.}. The quantum circuits used in these simulations went up to $4$ Trotter steps, reaching two-qubit gate depths of $146$ with a total of $5,034$ two-qubit gates. Note that some qubits near the center left of the chip suffer from large gate errors, and the electric energy greatly differs from the MPS simulation. However, away from this part of the lattice, the calculation on the quantum computer is able to reproduce the flow of electric energy away from the string. Additionally, due to the all-zero state not being an exact eigenstate of the theory, there are some fluctuations in the electric energy away from the string. 

As discussed in Section \ref{subsec:string_pheno}, adjoint string breaking in non-Abelian gauge theories displays a crossover between oscillation-dominated dynamics and string-breaking dominated dynamics. As a probe of this behavior, the operator $\hat{P}_0 \hat{P}_0$ was measured to determine the probability that the string breaks in the configuration shown in the top left of Fig.~\ref{fig:string_configs}. The upper and lower qubits were averaged over to reduce the statistical uncertainty. The operator $\hat{P}_1 \hat{P}_0$ was measured to determine the probability that the string oscillates to the configuration shown in the bottom left of Fig.~\ref{fig:string_configs}. Reflections across the $x$-axis were averaged over. Additionally, for the string oscillations, the probability for oscillating to the left or right was averaged over.

\begin{figure}
    \centering
    \includegraphics[width=\linewidth]{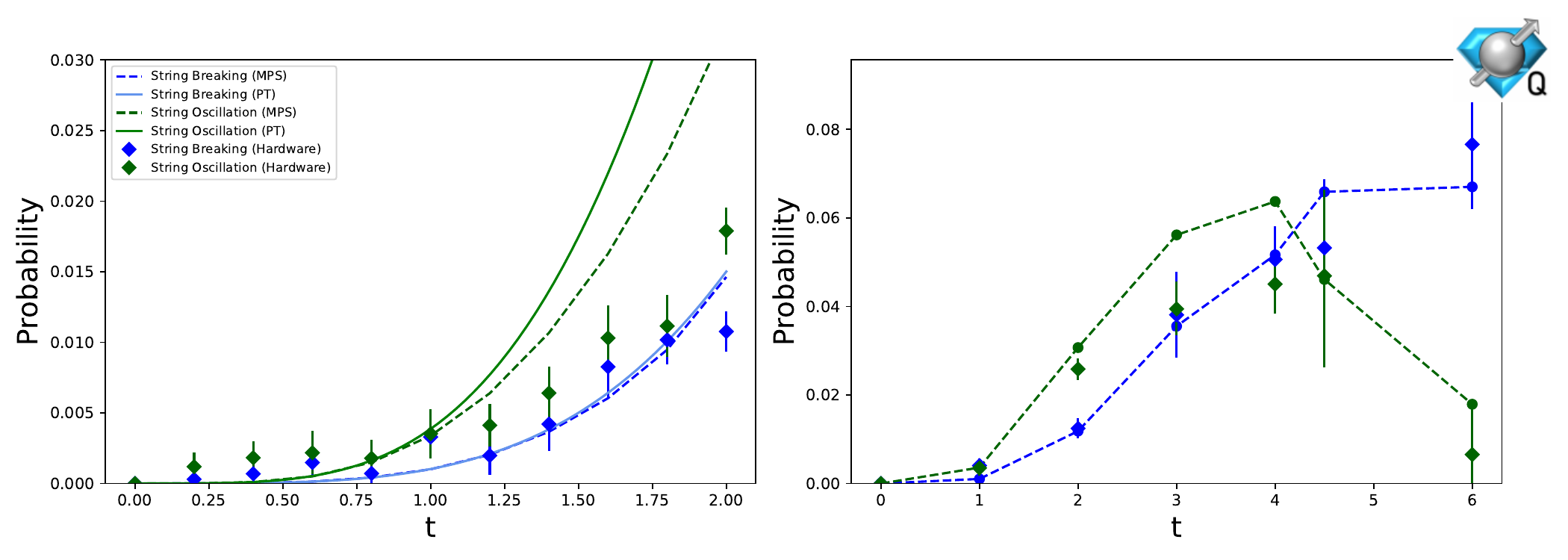}
    \caption{Probability of the adjoint string breaking and oscillating on an $8\times8$ lattice with $g=1.4$. The dashed curves are MPS calculations with a max bond dimension of 100. The data points are the mitigated results from {\tt ibm\_boston}. The solid lines are the time-dependent perturbation theory result computed in Appendix~\ref{sec:rates}. The left panel was computed using only two Trotter steps for each point and varying $\Delta t$. The right panel was computed using Trotter step sizes of $\Delta t = 0.5, 0.75, 1$. All simulations were performed using an even number of Trotter steps. For time slices sampled by multiple Trotter step sizes, only the results of the shallowest circuit are displayed.}
    \label{fig:break_oscillation.pdf}
\end{figure}

The left panel of Fig.~\ref{fig:break_oscillation.pdf} shows the evolution of the breaking and oscillation probabilities for short times using only two Trotter steps. Note that the left panel has no zero noise extrapolation applied (other than this, the default suite of error mitigation and suppression techniques discussed in Section \ref{sec:sim_method} are included). The right panel of Fig.~\ref{fig:break_oscillation.pdf} shows longer-time evolution using fixed Trotter step sizes of $\Delta t = 0.5, 0.75, 1$. All simulations were performed using an even number of Trotter steps. For time slices sampled by multiple Trotter step sizes, only the most shallow circuit was used. The deepest circuits used to generate this figure used $6$ Trotter steps and reached a two-qubit gate depth of $218$ with $7,634$ two-qubit gates present in the circuit. Only $15-50\%$ of the shots sampled survived the post-selection process at each time step. As described in Appendix~\ref{sec:Mitigation}, the post-selection procedure significantly improves upon the unmitigated results. This post-selection procedure is similar in spirit to the use of quantum error detection codes, which enable the identification of errors that can be removed through post-selection~\cite{Dasu:2026dwm,Froland:2026rzt,Froland:2026tfx}. However, our implementation requires no error-detecting code as our circuits to implement Trotterized time evolution have a built-in redundancy.

At short times, the MPS simulations are consistent with the time evolution predicted by perturbation theory. This is an indication that Trotterized time evolution is accurately reproducing the exact evolution at short evolution times. The hardware results' consistency with the classical simulations on these timescales indicates the hardware dynamics are dominated by oscillations at short times as predicted. While the hardware results do not quite reproduce the tensor network simulation at all times, the hardware does demonstrate a crossover between oscillations and breaking, validating the non-Abelian nature of the string dynamics in this simulation. Note that in these simulations, the breaking probability is reproduced with a higher fidelity than the oscillation probability. This is due to the breaking probability coming from an observable defined on neighboring qubits. The qubits used to compute the oscillation probability are spatially separated, leading to the probability being sensitive to any errors that occur in the path between the two qubits. 

\subsection{Dense Encoding}

\begin{figure}
    \centering
    \includegraphics[width=\linewidth]{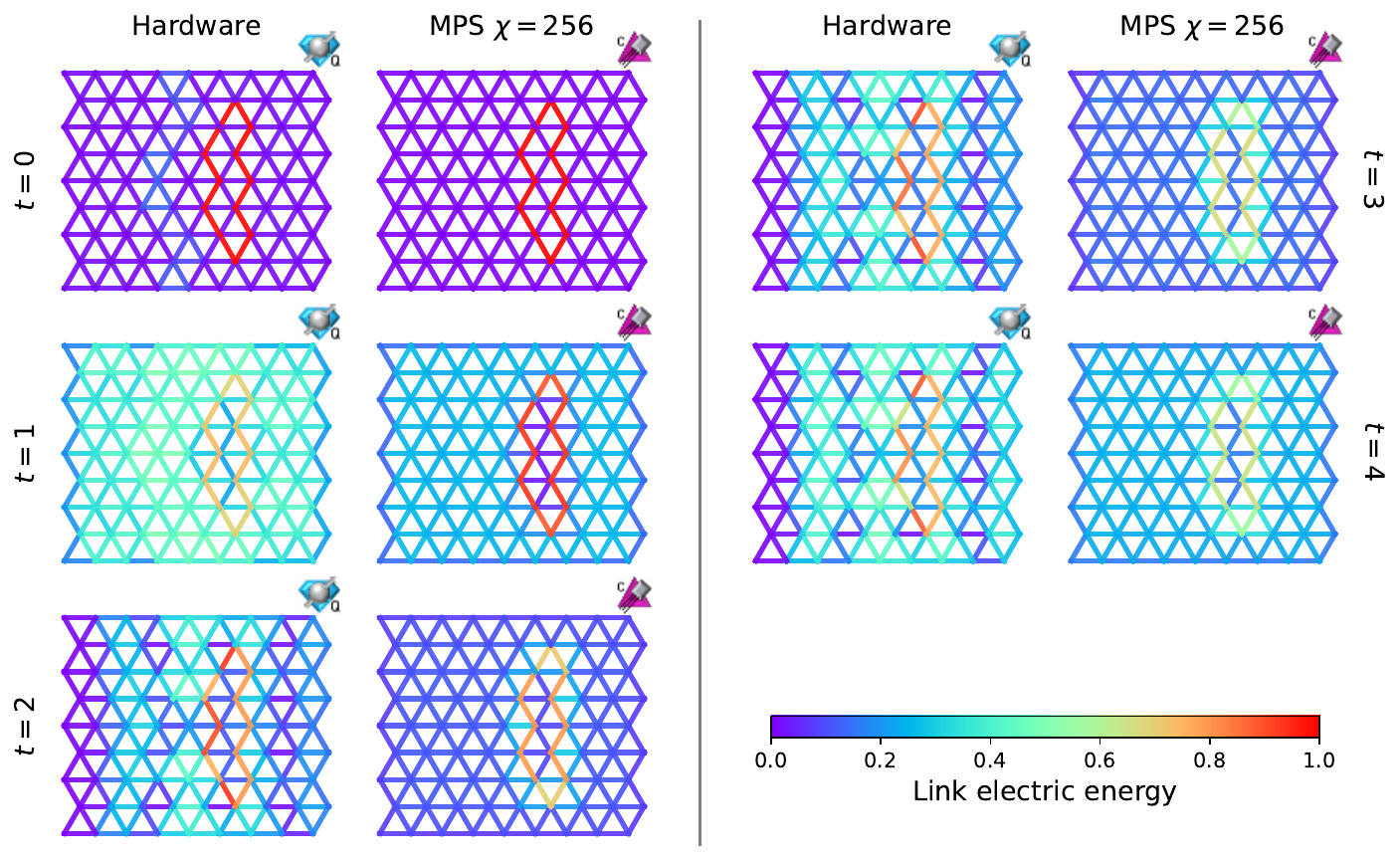}
    \caption{Evolution of the adjoint string on a $16\times8$ lattice. The Trotterized time evolution operator was synthesized on the dense encoding using BQSKit with $g=1.4$ and $\Delta t = 0.5$. The hardware results were computed on {\tt ibm\_boston} without applying ODR.
    }
    \label{fig:DenseEvolutionG14}
\end{figure}

We next describe quantum simulations on a $16 \times 8$ lattice mapped to heavy-hex following the dense encoding. Simulations were performed on this lattice with $g=1.4$ and $\Delta t = 0.5$. The results of these simulations are shown in Fig.~\ref{fig:DenseEvolutionG14}.
Note that by $t=4$ (8 Trotter steps), the circuit reaches a CZ depth of $577$ containing a total of $16,134$ CZ gates. Despite the large circuit depth in this simulation, the string is still visible in the hardware results, and the hardware produces time evolution that is qualitatively consistent with the result of tensor network simulations.

With regards to error mitigation using operator decoherence renormalization (ODR, \cite{Farrell:2023fgd}), note that approximating the densely encoded circuit by a Clifford circuit does not correspond to setting $\Delta t = 0$, and instead gives a Clifford circuit where the expectation value of most $\hat{Z}$ operators is $0$. This limits the applicability of ODR, which therefore we do not use for this encoding. Moreover, due to the different hardware mapping, we cannot take advantage of ancilla qubits to detect errors. ZNE also failed due to the larger circuit depths present in these circuits. Beyond these exceptions, we apply all the same error suppression/mitigation techniques as for the geometric encoding in Section~\ref{subsec:geo_encoding}.

There is a clear tradeoff between the number of sites that can be simulated and the circuit depth needed when comparing between the geometric and dense encodings. The $16 \times 8$ lattice simulated by the dense encoding, using all 156 qubits, is twice as large as the largest triangular lattice that can be simulated with the geometric encoding ($8 \times 8$, using 148 qubits). This larger lattice size enabled by the dense encoding reduces the potential adverse impact of the lattice boundaries on the string dynamics. On the other hand, the two-qubit gate depth required for carrying out e.g.~8 Trotter steps is almost twice as large for the dense encoding, 577 vs.~290.

\FloatBarrier
\section{Intermediate Coupling Regime}

Performing quantum simulations of lattice gauge theories at large gauge coupling values enables the clear identification of different types of dynamics and avoids the need to prepare potentially complicated initial states. However, the continuum limit lies at $ g\rightarrow 0$. As this limit is approached, computational basis states become high-energy states dominated by lattice artifacts. Therefore, it is necessary to prepare low-energy states to extract meaningful physics. As a step towards this limit, simulations were performed on an $8\times8$ lattice with $g=1.1$ using the geometric encoding. At this value of the coupling, the electric vacuum is no longer a good approximation of the vacuum state. An approximation to the vacuum state was prepared variationally using the ansatz
\begin{equation}
    \label{eq:vac_ansatz}
    \ket{\psi(\theta)} = \left(\prod_p e^{-i \theta \hat{Y}_p}\right) \ket{0} \ \ \ .
\end{equation}
$\theta$ was found by minimizing the expectation of the Hamiltonian. On a $4\times4$ lattice, it was found that this ansatz produced a state whose overlap with the actual vacuum is $\approx 0.987$. An adjoint string was placed in the same position as in the previous section by applying a Pauli $\hat{X}$ operator to all qubits within the string. This is not exactly the same as applying the string operator defined in Eq.~\ref{eq:string_creation_op}, but at intermediate values of $g$ produces a similar state with lower circuit costs. Note that at this value of the coupling, the vacuum expectation of the pairs of projectors used to track string breaking and oscillations is small, enabling their use to track dynamics at this coupling.

\begin{figure}
    \centering
    \includegraphics[width=0.5\linewidth]{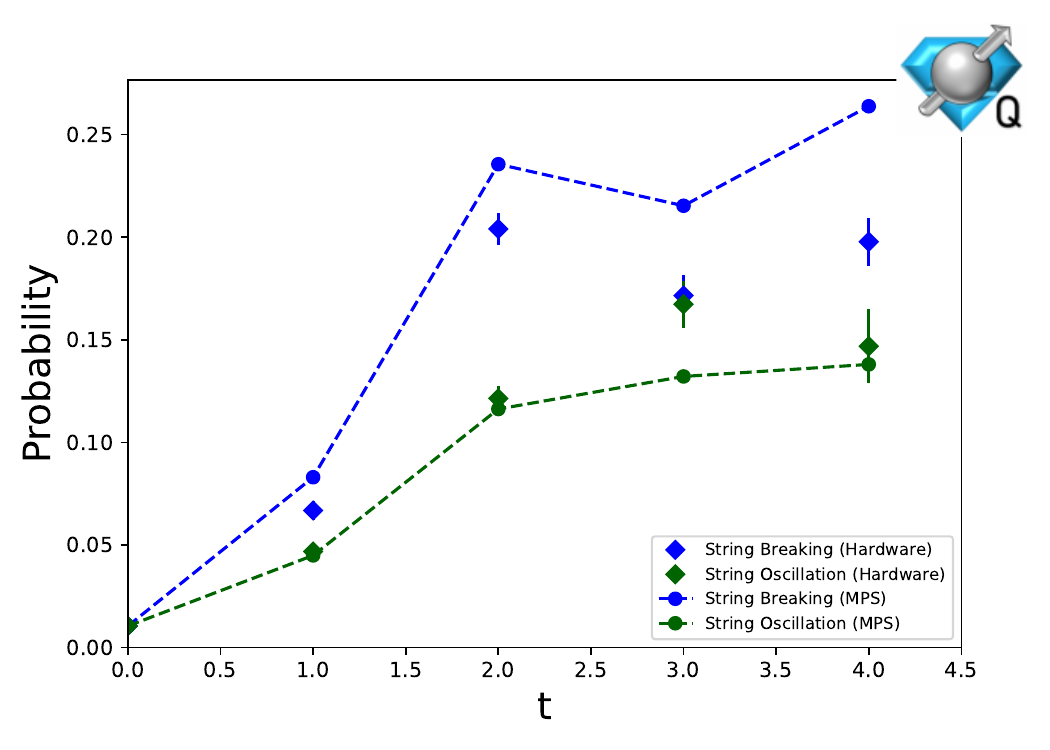}
    \caption{Probability of the adjoint string breaking and oscillating on an $8\times8$ lattice with $g=1.1$. The dashed curves are MPS calculations with a max bond dimension of 100. The solid curves are the mitigated results from {\tt ibm\_boston}.}
    \label{fig:break_oscillation11}
\end{figure}

For the quantum hardware calculations in this section, we use all of the error suppression and mitigation techniques described in Section~\ref{sec:sim_method}. Fig.~\ref{fig:break_oscillation11} shows the probabilities for the string to break or oscillate in the same manner as the previous section as a function of time. In these simulations, Trotter step sizes of $\Delta t = 0.5$, $\Delta t = 0.75$, and $\Delta t = 1$ were used. For time slices sampled by multiple Trotter step sizes, only the results of the shallowest circuit at that time step were used. While lacking precise agreement, these simulations demonstrate that the dynamics of the string are dominated by breaking processes at this coupling. Note that at this value of the coupling, the dynamics are faster than at $g=1.4$. The Trotter step sizes and circuit depths used in these calculations are the same as in the $g=1.4$ geometric encoding case.

\begin{figure}
    \centering
    \includegraphics[width=\linewidth]{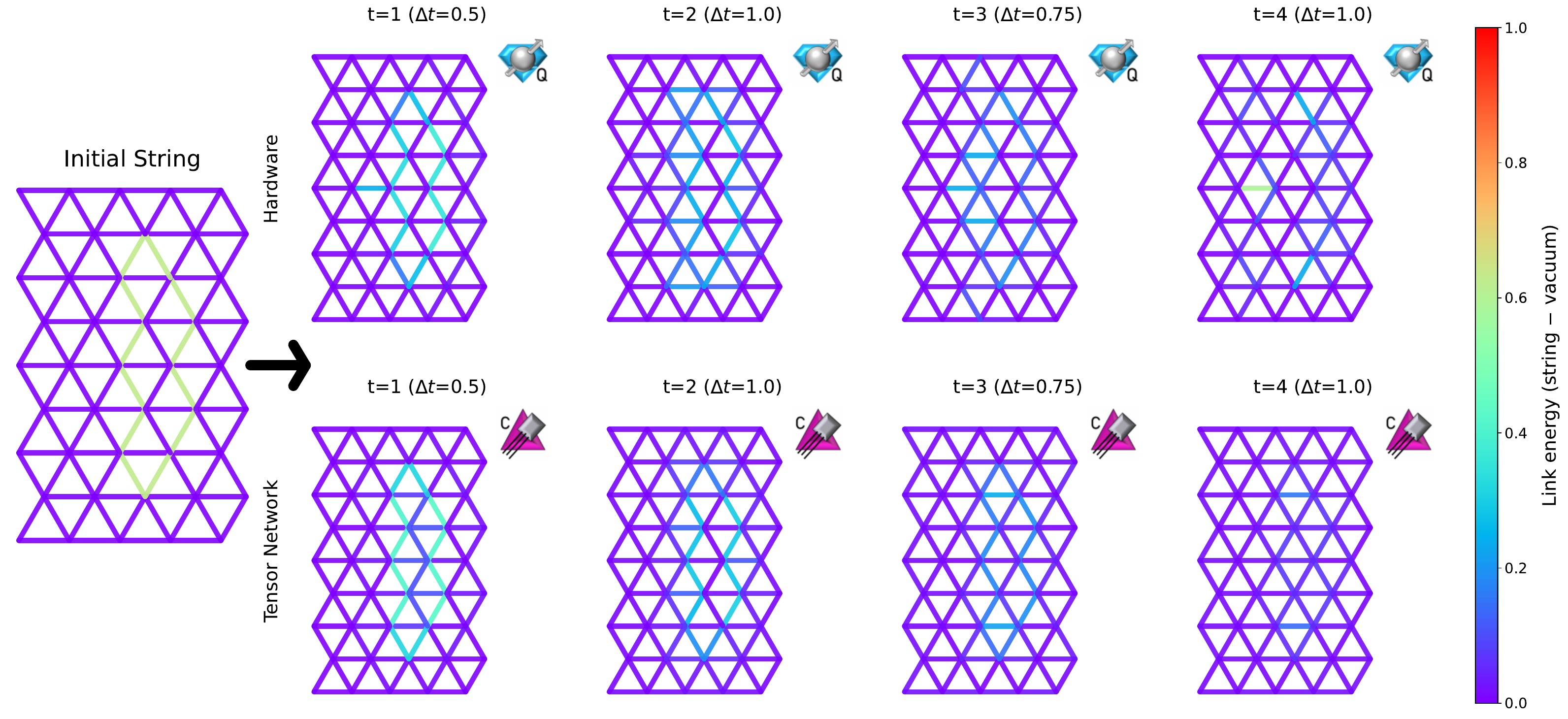}
    \caption{Evolution of an adjoint string on an $8\times8$ lattice with $g=1.1$. Each link shows the difference in the link energy between the state with the string and the vacuum state. The upper plots show the error-mitigated results obtained from {\tt ibm\_boston} and the lower plots show the results of an MPS simulation with max bond dimension $100$.
    The fading out of the electric energy along the string reflects the growing superposition of different string configurations and a shift of the energy from the electric term to the magnetic term.}
    \label{fig:string_ev_8x8_g11}
\end{figure}

Fig.~\ref{fig:string_ev_8x8_g11} shows the evolution of the electric energy on each link. To highlight the signal due to the string itself, and as an additional form of error mitigation, we have subtracted the electric energy for the evolved vacuum, where the vacuum is approximated following Eq.~(\ref{eq:vac_ansatz}). The difference between a linear and quadratic noise extrapolation in the context of ZNE was comparable in size to the link energies for several links, so the expectation of the link energy was only mitigated using ODR and post-selection, not ZNE.
At this value of the coupling, there is a rapid decrease in the electric energy along the string, while the probability of the string breaking or oscillating grows. This decrease in electric energy along the string is necessary due to energy conservation. The total electric energy of the state is proportional to the length of the string and limits what configurations can be created by the dynamics. The plaquette term couples different string configurations and therefore increases as the state evolves to a greater superposition of string configurations. Therefore, the decrease in the electric energy observed in this simulation can be interpreted as a shift of the energy from the electric term in the Hamiltonian to the magnetic term.

\section{Discussion}

The dynamics of adjoint strings are directly relevant to the hadronization of gluon jets produced in high-energy collisions. In current Monte Carlo event generators, gluons are modeled as effective color–anticolor pairs, so that adjoint string breaking is subsumed into fundamental string fragmentation and is never directly modeled~\cite{Sjostrand:2006za}. This neglects the possibility of gluon-mediated screening, which is the only source of string breaking in this work. The quantum simulations presented here are a first step towards directly making predictions about the hadronization of gluon jets from QCD. Explicitly, the simulations performed in this work demonstrate that the early-time dynamics of string breaking in two-dimensional non-Abelian lattice gauge theories are accessible to existing quantum computers. 
These simulations were made possible by the Krylov-based truncation of the gauge fields, which preserves locality and the symmetries of the Hamiltonian while enabling efficient mappings onto hardware.
The rapid convergence of Krylov-based truncations to the untruncated theory~\cite{Ciavarella:2025tdl,Modi:2026syn} provides confidence that results obtained at this truncation level can be systematically improved.
The use of BQSKit's quantum compilation enabled quantum simulations to be performed on hardware at previously inaccessible system sizes, allowing the simulation of electric link energies on lattices of up to $16 \times 8 = 128$ plaquettes using all 156 qubits of {\tt ibm\_boston}. 

Adjoint string dynamics has a crossover between oscillation-dominated behavior at early times and breaking-dominated behavior at long times. 
This crossover is a direct consequence of the non-Abelian nature of the SU(2) gauge group.
The quantum simulations on {\tt ibm\_boston} reproduce this crossover qualitatively at $g = 1.4$, demonstrating that the non-Abelian structure of the gauge group leaves a detectable imprint on the dynamics even in the presence of hardware noise. 
This constitutes a non-trivial validation that the quantum computer is faithfully representing the underlying gauge theory rather than an effectively Abelian substitute. At $g = 1.1$, the dynamics are faster and dominated more strongly by breaking, consistent with the expectation that moving toward the continuum limit suppresses the strong-coupling resonance structure that sustains coherent oscillations.

It is anticipated that the real-time dynamics of hadronization are classically hard to simulate in multiple spatial dimensions. The timescales simulated on quantum hardware in this work are still accessible to tensor network methods. 
Our classical simulations show that tensor networks can accurately reproduce our results from the quantum computer with only a few minutes of runtime. However, extending the simulation to longer timescales with PEPS requires more memory than a single Perlmutter node has available. MPS simulations are able to converge while being run on a single node, but require multiple days of runtime to simulate $40$ Trotter steps. $g=1.4$ is in the strong coupling regime, where the dynamics should be relatively weakly entangling and easy to compute with classical resources. As simulations move toward weaker coupling and more strongly entangled dynamics, classical tensor network methods will face even steeper computational barriers, further motivating the use of quantum hardware to access these regimes. Establishing a direct connection between first-principles lattice simulations and these collider observables would require performing simulations with fermions present, an SU(3) gauge group, in three spatial dimensions, and a controlled approach to the continuum limit. The theoretical framework to extend the truncation used in this work to include all these extensions exists~\cite{Ciavarella:2025bsg,Modi:2026syn}, and the developments in this work provide an important step towards implementation.

Several natural extensions of this work will be important for realizing its full potential. The $(1,2,1)$ truncation used here retains only the lowest-lying electric basis states. Extending to higher truncation levels will require more qubits but will enable a more faithful representation of the full theory and allow for a systematic extrapolation toward the continuum limit. Further integration of advanced compilation tools will be important for compiling the time evolution operator at these higher truncations. 
Finally, as hardware capabilities improve and error rates decrease, it will become possible to simulate larger systems, access longer evolution times, and employ more sophisticated error mitigation and error correction strategies, with the detection and post-selection techniques used in this work serving as a natural foundation for more powerful fault-tolerant protocols. 
In summary, this work establishes the viability of quantum simulation as a tool for probing two-dimensional non-Abelian gauge dynamics, demonstrates a qualitatively new type of string dynamics accessible only in higher dimensions and non-Abelian theories, and lays the groundwork for a systematic program of first-principles quantum simulation of hadronization physics.

\begin{acknowledgements}
    The authors would like to acknowledge helpful conversations with Christian Bauer, Ivan Burbano, Jesse Stryker, Irian D'Andrea, Neel Modi, Jad Halimeh, and Anupam Mitra. A.N.C was supported by the US Department of Energy, Office of Science, National Quantum Information Science Research Centers, Quantum Systems Accelerator (Award No. DE-SCL0000121). This research used resources of the National Energy Research Scientific Computing Center (NERSC), a Department of Energy User Facility under Contract No. DE-AC02-05CH11231 using NERSC award DDR-ERCAP0035884 and DDR-ERCAP0038362. ER is supported by the U.S. Department of Energy (DOE) under Contract No. DE-AC02-05CH11231, through the National Energy Research Scientific Computing Center (NERSC), an Office of Science User Facility located at Lawrence Berkeley National Laboratory. EY acknowledges support from the U.S. Department of Energy (DOE) under Contract No. DE-AC02-05CH11231, through the Office of Advanced Scientific Computing Research Accelerated Research for Quantum Computing Program, MACH-Q project. We acknowledge IBM’s Quantum Algorithm Engineering team for their insights and contributions. The authors acknowledge the use of Claude Opus 4.8 and 5 in the formatting of the plots in this paper. The data that supports the findings of this article is publicly available~\cite{su2data}.
\end{acknowledgements}

\appendix

\section{Truncated Hamiltonian}
\label{sec:Hamiltonian}
The local Krylov basis truncation works by generating basis states from repeated applications of operators in the Hamiltonian to a reference state~\cite{Ciavarella:2025bsg,Modi:2026syn}. This method can be used to construct an electric truncation by applying plaquette operators to the electric vacuum. Explicitly, the allowed states have nonzero overlap with states of the form
\begin{equation}
    \ket{\Vec{n}} = \prod_p \hat{\Box}^{\dagger \ n_p}_p \ket{0}\, ,    
\end{equation}
where the product is over all plaquettes on the lattice and $\ket{0}$ is the electric vacuum state.
Each application of the plaquette operators creates new loops of electric flux, and one can raise the electric truncation by allowing repeated applications of the same plaquette operator. This approach gives electric basis truncations defined by three integers, $(n_P,n_L,k)$ where $n_P$ is the maximum number of times an individual plaquette operator can be applied, $n_L$ is the maximum number of times a specific link operator can be applied, and $k$ is a cutoff on the irreps allowed on each link. In this work, the $(1,2,1)$ truncation was used with an SU($2$) gauge group. The irreps of SU($2$) are specified by half-integers $j$, and at this truncation, $j$ can take the values $0$ or $1/2$. The plaquette matrix elements for an SU($N_c$) gauge group have a scaling with $N_c$ that depends on the states of neighboring plaquettes~\cite{Ciavarella:2024fzw}. This can be used to define an additional truncation. In this work, any local field configuration that requires more than four $\mathcal{O}(1/N_c)$ transitions to be created will not be included. This is necessary for the qubit mapping used in this work, as otherwise, once the link irreps are fixed, there could be multiple ways to form a singlet state at a vertex. Using the mapping to qubits for the $(1,2,1)$ truncation and expressions for plaquette matrix elements from  Ref.~\cite{Ciavarella:2025bsg} gives the truncated Hamiltonian in Eq.~\ref{eq:SU2Ham}.
\section{String Breaking and Oscillation Rates}
\label{sec:rates}
In the strong coupling regime, the rate of string breaking and oscillations can be calculated using time-dependent perturbation theory. Explicitly, one can use an interaction picture where the electric piece of the Hamiltonian is taken to be the free part and the magnetic piece is the interaction. In this interaction picture, a plaquette operator at position $\Vec{r}$ is given by
\begin{align}
    \hat{\Box}_{\Vec{r},I}(t) & = \hat{P}^0_{\Vec{r} - \hat{x}} \hat{P}^0_{\Vec{r} + \hat{x}} \hat{P}^0_{\Vec{r} - \hat{y}} \left( \hat{b} e^{-i9/8g^2 t} + \hat{b}^\dagger e^{i9/8g^2 t} \right) \nonumber \\ 
    & + \frac{1}{2} \left(\hat{P}^1_{\Vec{r} - \hat{x}} \hat{P}^0_{\Vec{r} + \hat{x}} \hat{P}^0_{\Vec{r} - \hat{y}} + \hat{P}^0_{\Vec{r} - \hat{x}} \hat{P}^1_{\Vec{r} + \hat{x}} \hat{P}^0_{\Vec{r} - \hat{y}} + \hat{P}^0_{\Vec{r} - \hat{x}} \hat{P}^0_{\Vec{r} + \hat{x}} \hat{P}^1_{\Vec{r} - \hat{y}}\right) \left( \hat{b} e^{i3/8g^2 t} + \hat{b}^\dagger e^{-i3/8g^2 t} \right) \nonumber \\
    & + \frac{1}{4} \left(\hat{P}^0_{\Vec{r} - \hat{x}} \hat{P}^1_{\Vec{r} + \hat{x}} \hat{P}^1_{\Vec{r} - \hat{y}} + \hat{P}^1_{\Vec{r} - \hat{x}} \hat{P}^0_{\Vec{r} + \hat{x}} \hat{P}^1_{\Vec{r} - \hat{y}} + \hat{P}^1_{\Vec{r} - \hat{x}} \hat{P}^1_{\Vec{r} + \hat{x}} \hat{P}^0_{\Vec{r} - \hat{y}}\right) \left( \hat{b} e^{i3/8g^2 t} + \hat{b}^\dagger e^{-i3/8g^2 t} \right) \nonumber \\
    & + \frac{1}{8} \hat{P}^1_{\Vec{r} - \hat{x}} \hat{P}^1_{\Vec{r} + \hat{x}} \hat{P}^1_{\Vec{r} - \hat{y}} \left( \hat{b} e^{i9/8g^2 t} + \hat{b}^\dagger e^{-i9/8g^2 t} \right)\, ,
\end{align}
where $\hat{b}=\ket{0}\bra{1}$. The full interaction Hamiltonian is given by
\begin{equation}
    \hat{H}_I(t) = \frac{1}{g^2} \sum_{\Vec{r}} \hat{\Box}_{\Vec{r},I}(t) \ \ \ .
\end{equation}
The dominant time evolution behavior in the strong coupling regime will be set by second-order resonances. For two degenerate electric basis states ($\ket{i}$ and $\ket{f}$) with electric energy $\omega$, the matrix elements of the time evolution operator at this order in perturbation theory are
\begin{equation}
    \bra{f}\mathcal{T}e^{-i\int_0^t dt' \hat{H}_I(t')}\ket{i}_2 = -\sum_n \bra{f}\hat{H}_I(0)\ket{n}\bra{n}\hat{H}_I(0)\ket{i} \frac{t - \frac{e^{i(\omega - E_n)t}-1}{i(\omega - E_n)}}{i(E_n - \omega)} \ \ \ .
\end{equation}
As a concrete example, consider a $4\times2$ lattice with a string along the upper plaquettes. The amplitude for string breaking is given by
\begin{equation}
A_{B}(t) = \bra{\begin{matrix}
    1 &0 & 0 &1 \\ 0& 0 &0 &0
\end{matrix}}\mathcal{T}e^{-i\int_0^t dt' \hat{H}_I(t')}\ket{\begin{matrix}
    1 &1 & 1 &1 \\ 0& 0 &0 &0
\end{matrix}}_2 = \frac{2i}{3g^6} \left(t + \frac{e^{-i 3/8g^2 t} - 1}{i 3/8g^2}\right) \ \ \ .
\end{equation}
Similarly, the amplitude for the string to oscillate is given by
\begin{equation}
    A_{O}(t) = \bra{\begin{matrix}
    0 &1 & 1 &1 \\ 0& 1 &0 &0
\end{matrix}}\mathcal{T}e^{-i\int_0^t dt' \hat{H}_I(t')}\ket{\begin{matrix}
    1 &1 & 1 &1 \\ 0& 0 &0 &0
\end{matrix}}_2 = \frac{32}{9g^8} \left(1-\cos(3/8g^2 t)\right) \ \ \ .
\end{equation}
While $A_B(t)$ and $A_O(t)$ were computed for a short string on a small lattice, the same matrix elements are obtained for oscillations and breaking on longer strings on arbitrary lattices. Note that $A_O(t)$ does not contain a linearly growing piece due to cancellations in the sum over intermediate states. As a result, at long evolution times, the dynamics will be dominated by the string breaking. For small $t$, we find $
\abs{A_O(t)/A_B(t)} \approx 2$. Therefore, at short times the dynamics will be dominated by string oscillations, and at longer times there will be a transition to string breaking dominating the oscillations.

$A_B(t)$ and $A_O(t)$ give the amplitudes for specific string configurations. This is a global observable and sensitive to noise on the quantum computer. One can construct local observables with expectation values equal to $\abs{A_B(t)}^2$ and $\abs{A_O(t)}^2$ by noting that at 2nd order in perturbation theory $\mathcal{T}e^{-i\int_0^t dt' \hat{H}_I(t')}\ket{i}$ only has support on $\ket{i}$ and electric basis states where at most two qubits have flipped from the $\ket{i}$ state. Therefore, by measuring projectors on pairs of qubits, it is possible to project out each of these basis states and extract the probability of breaking or oscillation. These are the two-qubit operators that were measured in the main text.

\begin{figure}
    \centering
    \includegraphics[width=\linewidth]{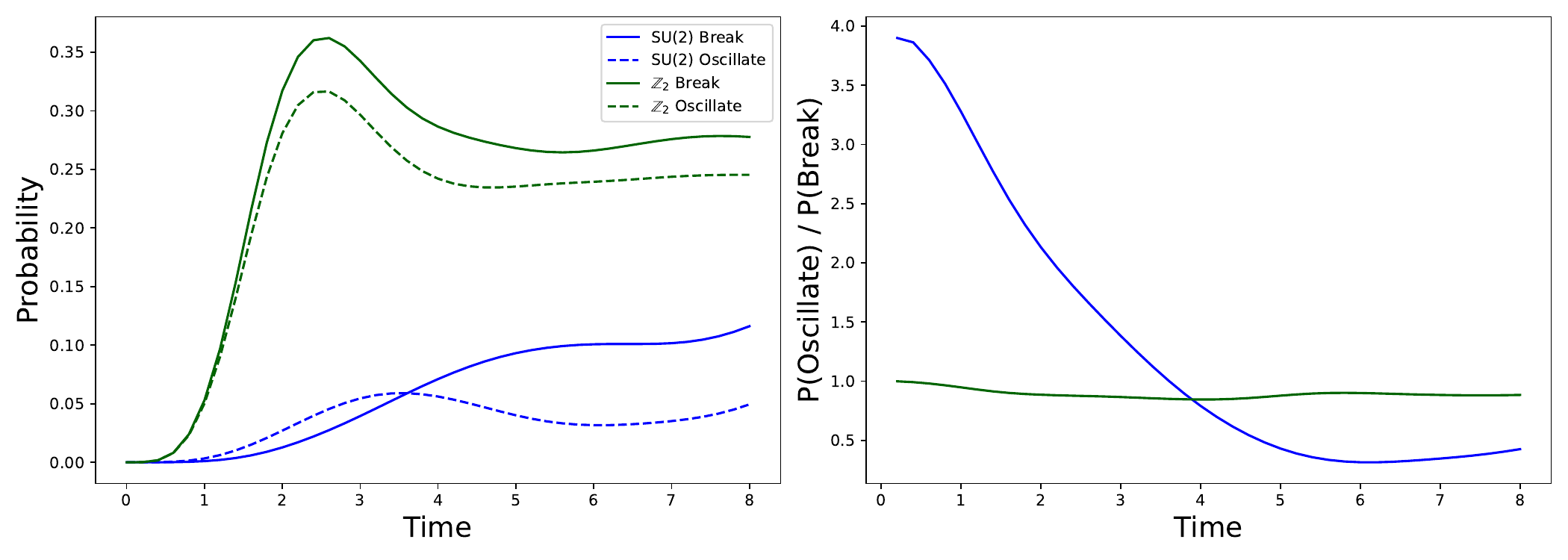}
    \caption{Breaking and oscillation probabilities for a string in a $\mathbb{Z}_2$ LGT and an SU(2) LGT. The left panel shows the evolution of the breaking and oscillation probabilities for both an SU(2) and $\mathbb{Z}_2$ lattice gauge theory at $g=1.4$. The right panel shows the ratio of the oscillation probability to the breaking probability as a funtion of time.}
    \label{fig:z2compare}
\end{figure}

It is interesting to compare the string breaking dynamics in an SU($2$) LGT to an abelian theory such as $\mathbb{Z}_2$. A $\mathbb{Z}_2$ LGT will have a Hamiltonian identical to the SU(2) Hamiltonian used in this work, except the plaquette operator is simply a $\hat{X}$ operator with no sum over projectors on the neighboring qubits. In other words, the matrix elements of the plaquette operator in a non-Abelian theory have a dependence on the electric flux flowing into the plaquette, which is not present in Abelian theories. As a consequence, in a $\mathbb{Z}_2$ theory, one would have $\abs{A_O(t)/A_B(t)} \approx 1$, and this switchover would not occur. Therefore, this dynamical crossover is a signature of the non-Abelian nature of the gauge group. This was validated numerically by performing an MPS simulation for $\mathbb{Z}_2$ LGT of an $8\times8$ lattice with $g=1.4$ and a string in the same position as in Section~\ref{sec:strongsim}. The simulation was run with the Trotterization used in Appendix~\ref{sec:ManualTrotter} and a step size of $\Delta t = 0.1$. Fig.~\ref{fig:z2compare} shows the probability of the string breaking or oscillating for both the SU(2) and $\mathbb{Z}_2$ lattice gauge theories. In the Abelian case, the breaking and oscillation probabilities are equal for short times, and in the non-Abelian case, there is a crossover between oscillation-dominated and breaking-dominated dynamics.


\section{Electric Truncation Error Calculation}
\label{sec:truncation}

The Hamiltonian studied in this work contains a truncation of the electric field, which will induce deviations in the dynamics from the full untruncated theory. As shown in previous work, the presence of Hilbert space fragmentation in the Kogut Susskind Hamiltonian leads to factorial convergence with the truncation of the electric field~\cite{Ciavarella:2025tdl}. The leading errors due to the field truncation can be estimated using time-dependent perturbation theory. Explicitly, one splits the Hamiltonian in the form $\hat{H} = \hat{H}_\Lambda + \hat{V}$ where $\hat{H}_\Lambda$ is the Hamiltonian with the plaquette term restricted to only have support on states where all links carry electric energy below $\Lambda$. For the truncation in this work, $\hat{H}_\Lambda$ would be our truncated Hamiltonian, and $\hat{V}$ would contain all plaquette matrix elements that couple to higher irrep states. As shown in Ref~\cite{Ciavarella:2025tdl}, the leading order expression for the probability that a link $l$ leaks out of the truncated Hilbert space by time $t$ is upper bounded by
\begin{equation}
    \epsilon_{j=1} \leq \text{max}_{B,N,N'} \abs{\int_0^t dt' e^{i t' \left(\bra{B, j_l = 1, N'}\hat{H}_\Lambda\ket{B, j_l = 1/2, N}-\bra{B, j_l = 1/2, N}\hat{H}_\Lambda\ket{B, j_l = 1/2, N}\right)} \bra{B,j_l=1,N'} \hat{V} \ket{B, j_l = 1/2, N}}^2 \, ,
\end{equation}
where $B$ corresponds to the state of links that don't share a plaquette with link $l$, $j_l$ is the irrep on link $l$, and $N$ corresponds to the links that lie on a plaquette that contains $l$.

For the states considered in this work, the dominant channel for leaving the truncated Hilbert space occurs when a plaquette next to the initial string is excited, and the shared link has a representation of $j=1$. The plaquette matrix element for this transition is given by $\sqrt{2/3}$~\cite{Ciavarella:2025bsg}, and using the formalism developed in previous work, the perturbative estimate for the probability that a given plaquette makes this transition is bounded by
\begin{equation}
    \epsilon_{j=1} \leq \frac{512}{\ \ 363g^8} \ \ \ .
\end{equation}
With $g=1.4$, we have $\epsilon_{j=1} \leq 0.095$ and with $g=1.1$ we have $\epsilon_{j=1} \leq 0.66$. Based on these estimates, we can conclude that the gauge field truncation errors are under control for the $g=1.4$ simulations, but $g=1.1$ would need a higher truncation to controllably approximate the dynamics of the untruncated theory.

\section{Approximation to the Vacuum State}
\label{sec:VacPrep}
\begin{figure}
    \centering
    \includegraphics[width=0.5\linewidth]{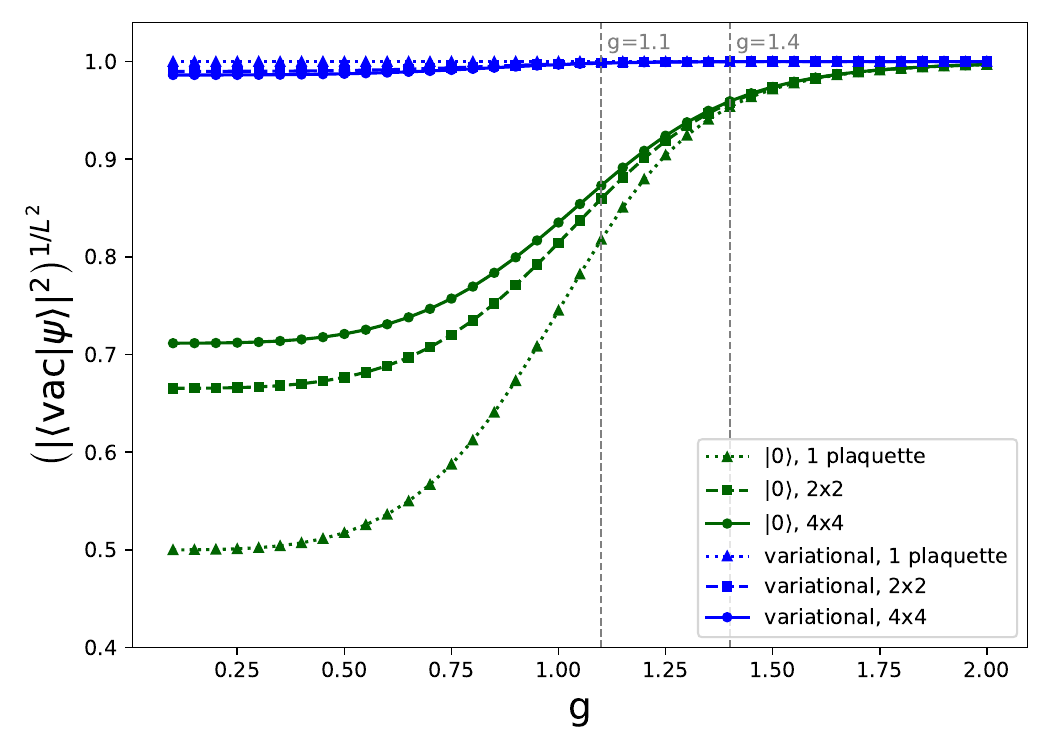}
    \caption{Overlap with the vacuum state at different system sizes as a function of $g$. The green curves show the overlap of the all-zero state with the vacuum of the system. The blue curves show the overlap of the true vacuum with the variational state given by Eq.~(\ref{eq:vac_ansatz}) found through minimizing the energy.}
    \label{fig:VacPrep}
\end{figure}

The simulations performed in this work require preparing an approximation to the vacuum state before adding the adjoint string to the system. In the $g\rightarrow\infty$ limit, the vacuum state is given by the electric vacuum (all $0$) state. However, at finite values of $g$, the vacuum state will be a more generic superposition state. In practice, the vacuum state can be approximated variationally through minimizing the energy of a state prepared by an ansatz circuit. In this work, the ansatz
\begin{equation}
    \ket{\psi(\theta)} = \left(\prod_p e^{-i \theta \hat{Y}_p}\right) \ket{0} \ \ \ ,
\end{equation}
will be used to approximate the vacuum state. This ansatz was chosen as the vacuum state of this theory has short-range correlations and can be written with only real state-vector coefficients. Additionally, up to boundary effects, the state should be translationally invariant, which is why there is only one free angle. The overlap of the true vacuum with the state found variationally for an $L\times L$ lattice is shown in Fig.~\ref{fig:VacPrep}. Note that state overlap is an extensive measure, and it is expected that a state with good overlap with the vacuum state will have its overlap $\abs{\bra{\text{Vac}}\ket{\psi}}^2$ fall off exponentially with $L^2$. This is reflected in the overlap of the curves for different system sizes at large $g$. This figure demonstrates that the electric vacuum state works well as an approximation of the vacuum when $g$ is large, but it breaks down as $g\rightarrow0$. The variational ansatz is able to maintain a much larger overlap with the true vacuum state. At $g=1.4$, the electric vacuum has a large overlap with the true vacuum at small system sizes, and the overlap with the true vacuum has the predicted scaling with system size. This is why for the main $g=1.4$ results in this paper the variational ansatz was not needed. At $g=1.1$, on the other hand, it is necessary to use the variational state to maintain accuracy as system size is increased.

\section{Geometric Circuit Construction}
\label{sec:ManualTrotter}

\begin{figure}
    \centering
    \includegraphics[width=0.75\linewidth]{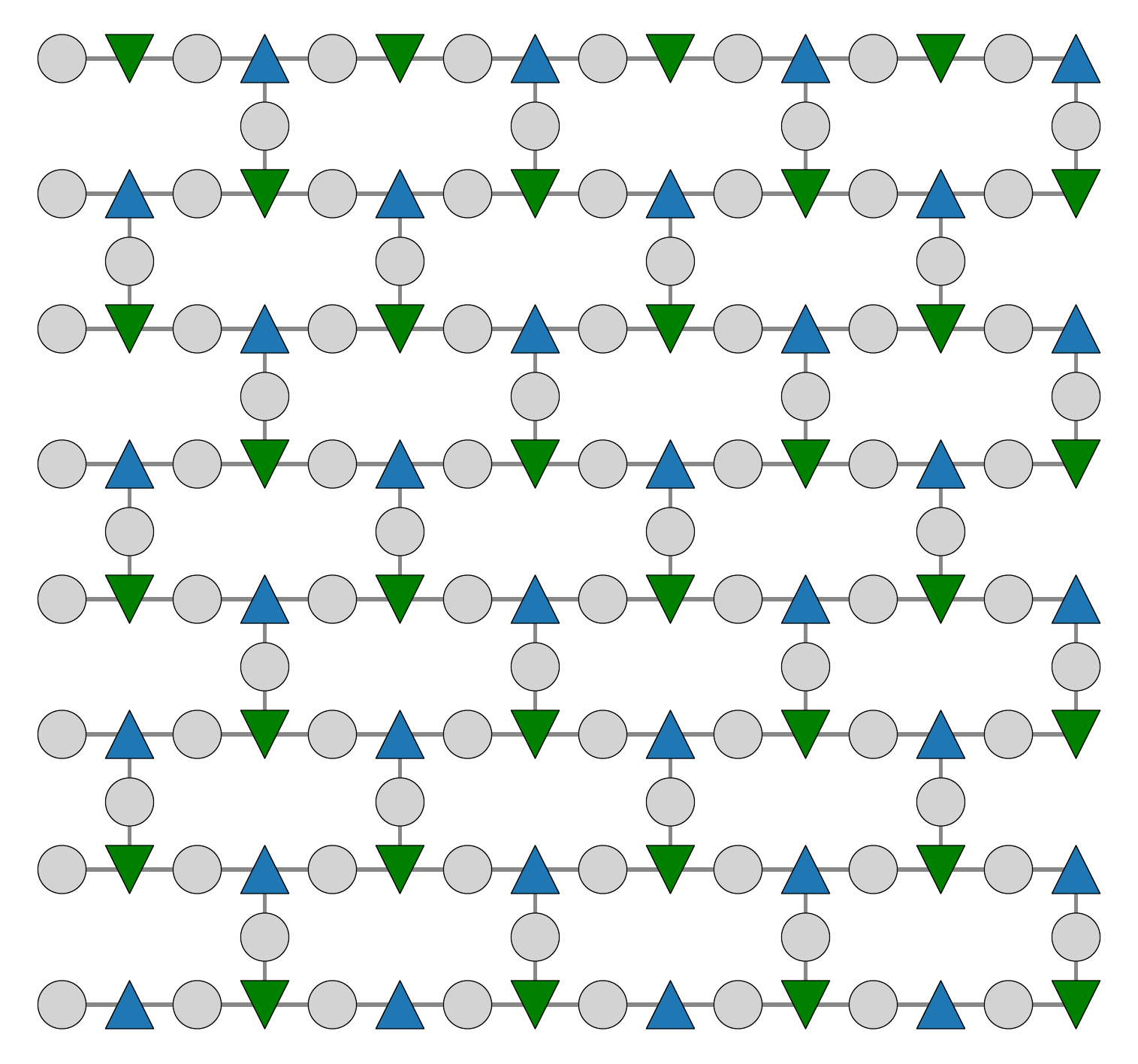}
    \caption{Encoding of the triangular lattice onto the heavy hex architecture. The blue and green triangles correspond to the plaquettes of the triangular lattice. Each of these triangles has a single qubit assigned to it. The grey circles correspond to ancilla qubits used to enable communication between the qubits that represent the state of the physical system being simulated.}
    \label{fig:HeronMapping}
\end{figure}

\begin{figure}
    \centering
    \includegraphics[width=\linewidth]{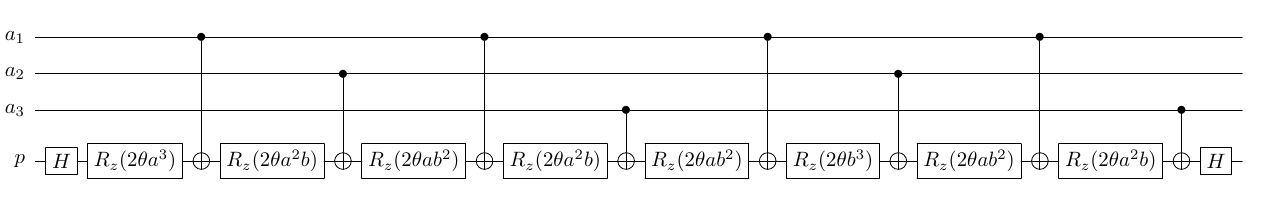}
    \caption{Circuit to implement $\exp{-i\theta \hat{C}_{a_1}\otimes \hat{C}_{a_2}\otimes\hat{C}_{a_3}\otimes \hat{X}_p }$ where $\hat{C}_q = a \hat{1} + b \hat{Z}_q$ and $R_z(\theta) = \exp{-i\theta \hat{X}}$. To implement $\exp{-i \theta \hat{\square}_p}$ with this circuit, one takes $a=\frac{3}{4}$, $b=\frac{1}{4}$.}
    \label{fig:plaq_circuit}
\end{figure}

A triangular lattice can be mapped onto the heavy hex architecture of the Heron processors as shown in Fig.\ref{fig:HeronMapping}, where the colored triangles represent the plaquettes and grey circular qubits are used as ancillas.
A Trotterized time evolution operator can be constructed by splitting the magnetic term into terms acting on two different sublattices. The even sublattice, $\mathcal{E}$, corresponds to all green qubits in Fig.~\ref{fig:HeronMapping} and the odd sublattice, $\mathcal{O}$, corresponds to all blue qubits. The even and odd magnetic Hamiltonians are given by
\begin{align}
    \hat{H}_{B,e} &= -\frac{1}{g^2} \sum_{p \in \mathcal{E}} \left(\prod_{\hat{n}} \hat{C}_{p+\hat{n}}\right) \hat{X}_p \nonumber \\
    \hat{H}_{B,o} &= -\frac{1}{g^2} \sum_{p \in \mathcal{O}} \left(\prod_{\hat{n}} \hat{C}_{p+\hat{n}}\right) \hat{X}_p \ \ \ .
\end{align}
With this decomposition of the magnetic term, time evolution was performed using the second-order Trotter decomposition,
\begin{equation}
    \hat{U}(\Delta t) = e^{-i \hat{H}_E \Delta t /2} e^{-i \hat{H}_{B,e} \Delta t /2} e^{-i \hat{H}_{B,o} \Delta t} e^{-i \hat{H}_{B,e} \Delta t /2} e^{-i \hat{H}_E \Delta t /2} \ \ \ .
\end{equation}
At the start of each Trotter step, the ancilla qubits are prepared by applying a CNOT gate to all ancillas, controlled by the neighboring blue qubit, thus entangling the ancillas with the blue qubit. 
The electric evolution operator, $e^{-i \hat{H}_E \Delta t /2}$, consists of rotations generated by single $\hat{Z}$ operators and $\hat{Z}\hat{Z}$ two qubit operators. The single $\hat{Z}$ rotations can be done using single-qubit gates, and the two-qubit $\hat{Z}\hat{Z}$ rotations can be implemented using a $2$ CNOT circuit acting on the green qubit and each of its neighboring ancillas. $e^{-i \hat{H}_{B,e} \Delta t /2}$ consists of $\hat{X}$ rotations on the green qubits controlled by all of the neighboring blue qubits. Since the state of all of the blue qubits has been ``copied'' onto the ancilla qubits, these rotations can be done controlled by the ancillas using the circuit shown in Figure
~\ref{fig:plaq_circuit}. After $e^{-i \hat{H}_{B,e} \Delta t /2}$ has been applied, CNOT gates are applied to all of the ancillas controlled by the neighboring blue qubits to return all ancillas to the zero state.  $e^{-i \hat{H}_{B,o} \Delta t /2}$ is applied in a similar manner, with the blue and green qubits switching roles. Note that the circuit for a single Trotter step begins and ends with CNOT gates being applied to the ancilla qubits and evolution under the electric Hamiltonian. This allows for the last layer of CNOTs in each Trotter step to cancel with the first layer of CNOTs in the following Trotter step.
Circuit transpilation in the end expresses the CNOT's in terms of controlled Z (CZ) gates, where each CNOT requires one CZ. The circuit depths and total number of CZ gates used in these circuits are given in Table~\ref{tab:GateCounts}
\begin{table}[]
    \centering
    \begin{tabular}{|c|c|c|}
    \hline
    Trotter Step Number & CZ Gate Count & CZ Depth  \\
    \hline
    2     & 2,434 & 74 \\
    \hline
    4     & 5,034 & 146 \\
    \hline
    6     & 7,634 & 218 \\
    \hline
    8     & 10,234 & 290 \\
    \hline
    \end{tabular}
    \caption{The CZ gate counts and CZ gate depths to implement different numbers of Trotter steps using the geometric circuit construction on an $8 \times 8$ lattice.}
    \label{tab:GateCounts}
\end{table}
\FloatBarrier

\section{Dense Circuit Construction}
\label{sec:tiledtrotter}

\begin{figure}
    \centering
    \includegraphics[width=\linewidth]{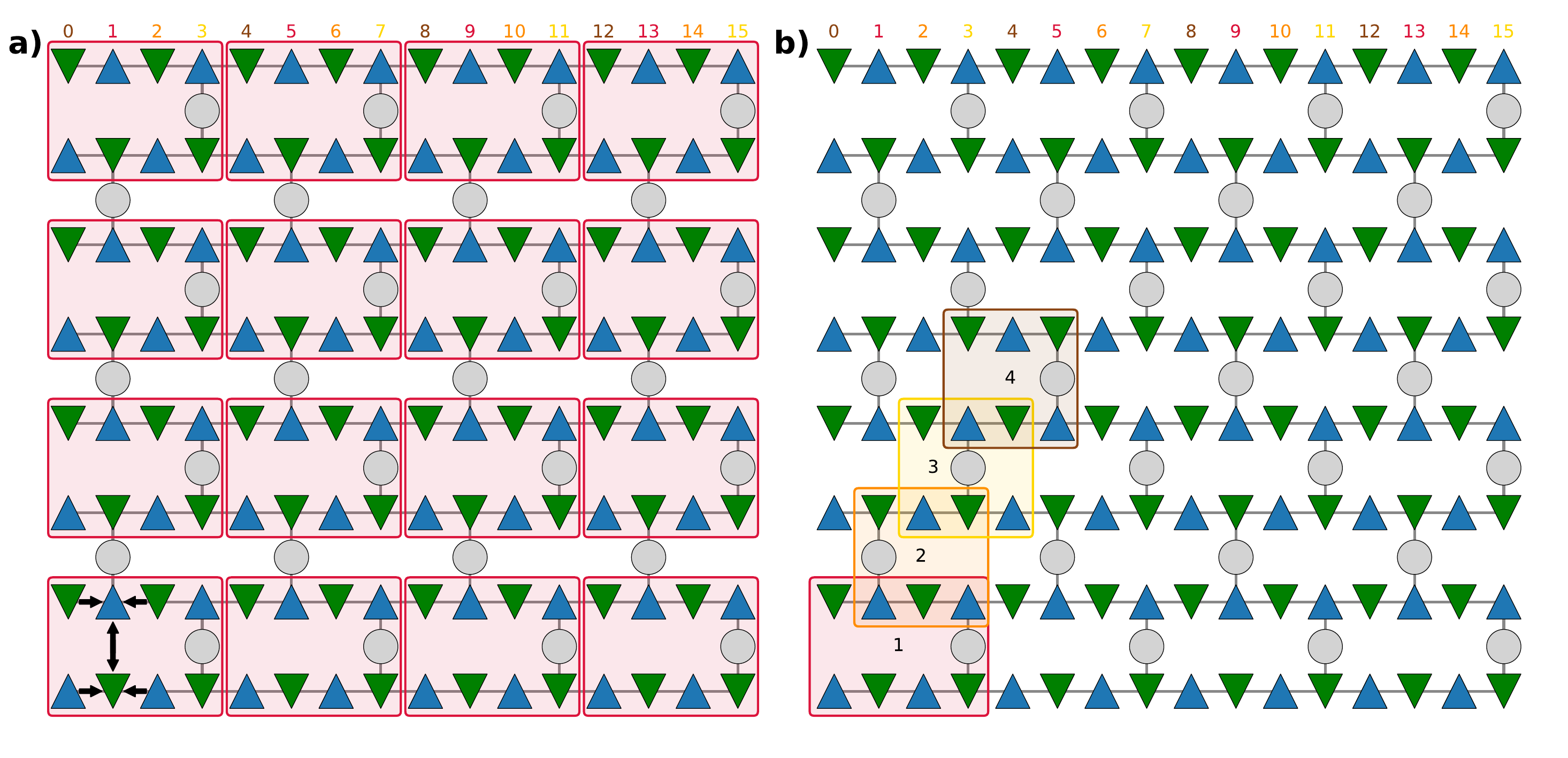}
    \caption{The dense encoding maps the lattice onto the hardware architecture via modular tiling and numerical synthesis. Triangles denote plaquettes; circles denote ancillas. Top labels indicate lattice columns. \textbf{(a)} A repeating $2\times4$ block tiles the chip. The bottom left block shows the required Hamiltonian interactions for column 1. BQSKit synthesizes these interactions from the complete unitary across the block's available linear connections, circumventing missing physical wiring. \textbf{(b)} The four tile variants needed for a complete Trotter step. Each tile represents one substep, color-coded and labeled to match its target column indices. When each tile is laid across the chip, it captures all the interactions in the global Hamiltonian for its target column. All four substeps together capture the global Hamiltonian.}
    \label{fig:DenseEncoding}
\end{figure}

Appendix \ref{sec:ManualTrotter} mapped the lattice onto the hardware, capping the maximum lattice size at $8\times8$. Here, we invert the approach: we map the hardware onto the lattice using numerical synthesis. This packs the chip efficiently, yielding a much denser $16\times8$ lattice. Figure~\ref{fig:DenseEncoding} illustrates this dense encoding.

We tile the chip with a repeating 9-qubit block arranged in a $4\times4$ grid. Each block maps to a $2\times4$ lattice section containing eight plaquette qubits and one ancilla qubit. To capture the global Hamiltonian, adjacent tiles must interact. We bridge these boundaries by slicing each Trotter evolution step into four substeps. With each substep, we shift the tiling grid across the hardware like a sliding window.

Visualize the lattice as vertical columns. Substep 1 captures the $H_E$ and $H_B$ interactions for the 1st column, modulo 4. Substep 2 shifts to the 2nd column, modulo 4. Substeps 3 and 4 sweep the remaining columns. Stacking these four shifts weaves the localized blocks together, capturing every term in the Hamiltonian. If a shifting tile pushes past the chip's edge, we simply truncate it.

We extract unitaries for these tiles directly from the Hamiltonian and synthesize them into quantum circuits using BQSKit~\cite{younis2021berkeley}. This approach overcomes hardware limits: even if the chip lacks specific physical connections between qubits, BQSKit synthesizes an executable circuit by efficiently routing operations over the available architecture within the tile. Notably, substeps 2, 3, and 4 demand the exact same unitary. However, because the underlying physical wiring varies across different regions of the chip, the compiler translates this single unitary into two distinct physical circuits. We build the full simulation step by dropping these synthesized sub-circuits onto the chip and layering the substeps.

This method carries a tradeoff: it hardcodes the parameters $g$ and $\Delta t$ into the physical circuits. Adjusting either parameter requires resynthesizing the tiles. Yet, this approach decouples the lattice size from the synthesis cost. Because the global circuit is built from localized, modular tiles, we can snap them together to construct varied lattice sizes, such as the $4\times8$ lattice used to benchmark the geometric circuits, without recompiling. This makes this use case of numerical synthesis highly scalable. When future hardware expands, we can simulate larger lattices simply by placing more tiles.

An additional tradeoff is in the depth of the circuits. The dense circuit construction enables simulation of a $2 \times$ larger physical lattice; however, this comes at the cost of about twice as deep circuits. The circuit depths and total number of CZ gates used in these circuits are given in Table~\ref{tab:TiledGateCounts}.

\begin{table}[]
    \centering
    \begin{tabular}{|c|c|c|}
    \hline
    Trotter Step Number & CZ Gate Count & CZ Depth  \\
    \hline
    2     & 4038 & 152 \\
    \hline
    4     & 8070 & 297 \\
    \hline
    6     & 12102 & 437 \\
    \hline
    8     & 16134 & 577 \\
    \hline
    \end{tabular}
    \caption{The CZ gate counts and CZ gate depths to implement different numbers of Trotter steps using the tiled circuit (dense) construction.}
    \label{tab:TiledGateCounts}
\end{table}
\FloatBarrier

\section{Error Mitigation}
\label{sec:Mitigation}
\begin{figure}
    \centering
    \includegraphics[width=\linewidth]{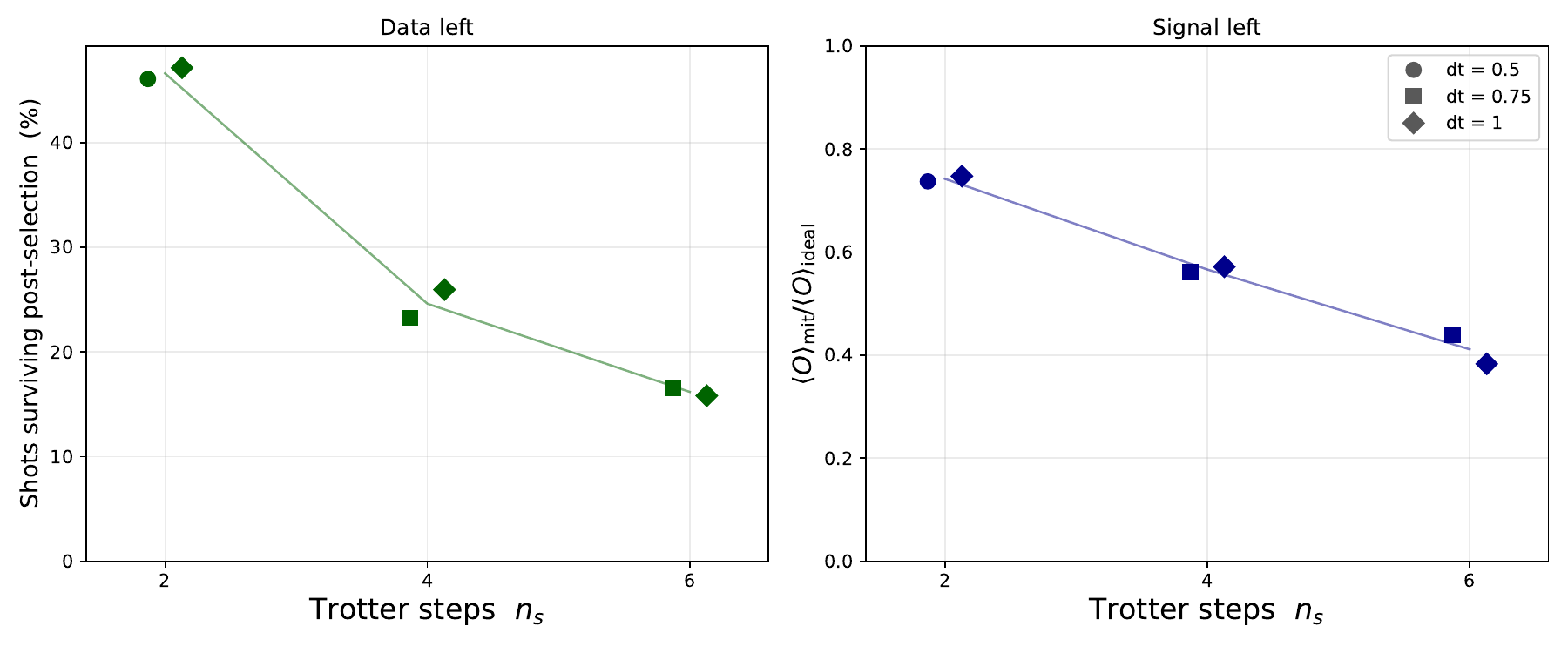}
    \caption{Percentage of shots surviving the post-selection and decoherence renormalization factor for the $ZZ$ operators measured in the string breaking probability for the $8\times8$ lattice at $g=1.4$ with no noise enhancement applied.}
    \label{fig:selectionODR}
\end{figure}

Existing quantum computers suffer from noise and errors that affect the quality of the results. In the simulations presented in this work, an XY4 dynamical decoupling sequence was used to reduce the size of coherent errors~\cite{Viola:1998gg}. Pauli twirling was used to convert all coherent errors induced by entangling gates into incoherent Pauli errors~\cite{Wallman:2015uzh}. Measurement errors were mitigated using twirled readout extinction of errors (TREX)~\cite{Berg:2020ibi}. Each twirl was sampled with $2000$ shots, and $64$ twirls were performed per circuit. All statistical errors were estimated using bootstrapping. By modeling the remaining noise as the exact quantum circuit followed by a Pauli error channel (assuming the Pauli error channel does not depend on the single-qubit rotation angles), we can mitigate the remaining noise using operator decoherence renormalization (ODR)~\cite{Urbanek:2021oej,ARahman:2022tkr,Farrell:2023fgd}. ODR works by running a mitigation circuit with the same two-qubit gate structure as the original circuit, with the single-qubit rotations modified to make the mitigation circuit classically simulable. In this work, the mitigation circuit was chosen by performing forward time evolution for the first half of the circuit and backwards evolution for the second half of the circuit. Note that this choice of mitigation circuit is restricted to using an even number of Trotter steps. Under the assumed noise model, the expectation of Pauli operators is rescaled by an operator-dependent factor, i.e.
\begin{equation}
    \langle O \rangle_N = \lambda_O \langle O \rangle_I \ \ \ , 
\end{equation}
where $\langle O \rangle_I$ is the result that would be obtained from an error-free quantum computer, and $\langle O \rangle_N$ is the result obtained from the noisy quantum computer. $\lambda_O$ is computed using the mitigation circuit, and then used to rescale the results from the circuit we wish to implement. The projectors computed in this work were decomposed into a sum over products of Pauli $Z$ operators, and ODR was applied to each $Z$-based Pauli string independently. The mitigated $Z$ expectation values were then added together to obtain the mitigated projector expectation values. This is the same procedure that was used in Ref.~\cite{Froland:2026aff}. The right panel of Fig.~\ref{fig:selectionODR} shows the average rescaling factor used for the $ZZ$ operators in the $g=1.4$ $8\times8$ string breaking probability as a function of Trotter step number.

On the actual quantum computer, not all noise matches the model assumed by ODR. This remaining noise was mitigated through a zero noise extrapolation (ZNE). This technique works by running additional versions of circuits where the noise has been artificially enhanced. In this work, the noise was enhanced by replacing each CNOT gate in the circuit with $3$ CNOT gates with probability $p$ (probabilistic gate folding). In the absence of device errors, this would not change the unitary implemented by the circuit. This was done with $p=0.25$ and $p=0.5$, leading to average depth enhancements of $r=1.5$ and $r=2$. ODR was performed for the original and noise-enhanced circuits. Then a linear extrapolation to $r=0$ was performed. The results of the linear extrapolation were compared to a quadratic extrapolation to confirm that the extrapolation was reliable.

\begin{figure}
    \centering
    \includegraphics[width=\linewidth]{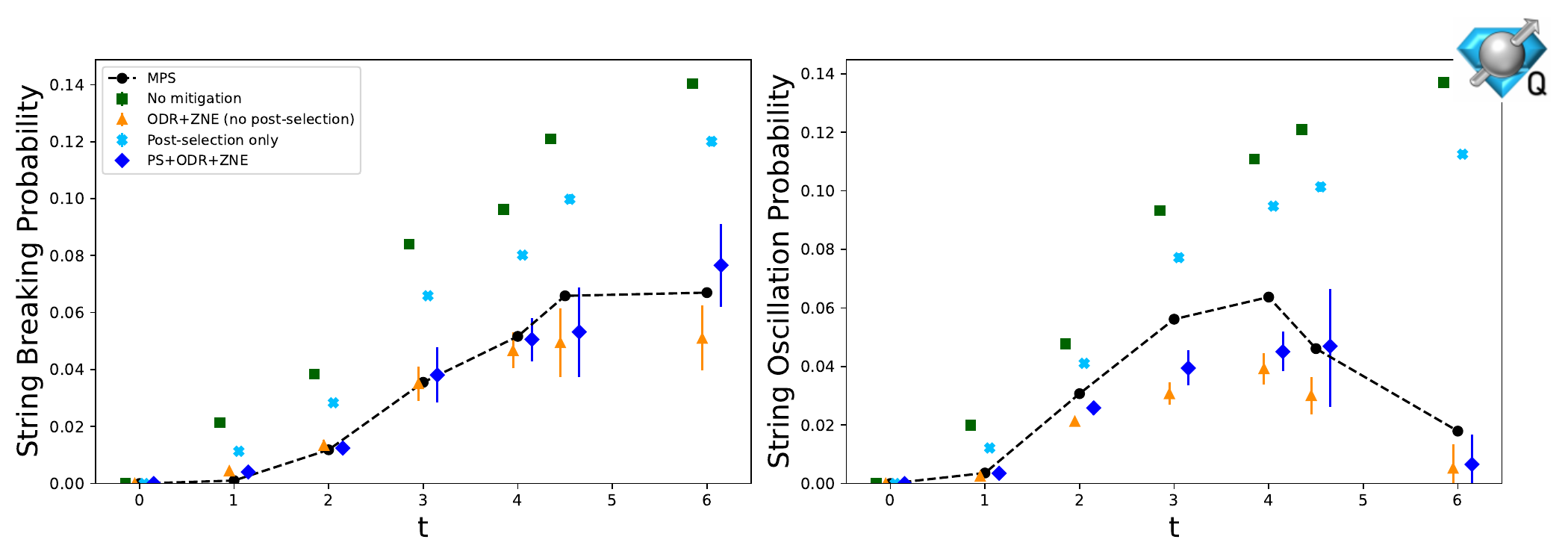}
    \caption{Effects of different error mitigation strategies on the breaking (left) and oscillation (right) probabilities computed on an $8\times8$ lattice with $g=1.4$. The green points are the raw results from the hardware with no error mitigation. The light blue points are the post-selected results. The blue and orange points have ODR and the zero-noise extrapolation applied. The black points (connected by interpolating dashed lines) were computed using MPS tensor networks.}
    \label{fig:MitigationComparison}
\end{figure}

The above error mitigation techniques are general and can, in principle, be applied to any quantum circuit. The circuits used to implement the time evolution operator make use of ancilla qubits that, at the end of each Trotter step, are returned to the $\ket{0}$ state. Due to errors in the hardware, the ancilla qubits do not always return to the $\ket{0}$ state. Therefore, one can post-select on the ancilla qubits to mitigate errors. Due to the large size of the lattice, post-selecting on all ancilla qubits would leave few remaining samples. All observables computed in this work are products of a small number of local operators. For a given local observable, a post-selection was performed based on the state of all of the ancilla qubits neighboring the physical qubits that the operator has support on. The percentage of shots that survive the post-selection in the computation of the string-breaking probability at $g=1.4$ is shown in Fig.~\ref{fig:selectionODR}. The effects of the different stages of error mitigation is shown in Fig.~\ref{fig:MitigationComparison}. It is interesting to note that even post-selection on its own noticeably shifts the results towards the correct values.

Additionally, all configurations in this work are symmetric about reflections across the x-axis. Therefore, one can average all expectation values with their reflected counterpart to reduce the effect of errors. This allows one to address an issue with ODR where a mitigated expectation value lies outside the physical range (i.e., reporting a mitigated probability being larger than one or being negative). If a mitigated observable lies outside the physical range, it is discarded, and the value of the reflected observable is reported for both. Otherwise, the observable and its reflected counterpart are averaged together.

\section{Computational Cost and Convergence of Classical Simulations}
\label{sec:TensorSim}
The quantum simulations performed in this work are still amenable to classical simulation. However, it is expected that the dynamics of hadronization become classically hard to simulate at larger timescales. To probe this, tensor network simulations using the Quimb tensor network library~\cite{Gray:2018psf} were performed on the $8\times8$ lattice with $g=1.4$ and $\Delta t= 0.5$. These simulations were run using a single CPU node on the Perlmutter compute cluster. Simulations were performed using both MPS and PEPS tensor networks. Each simulation was allowed to use a total of $5$ threads, enabling different bond dimension simulations to be run in parallel. Gates in the PEPS simulations were applied using a simple update with a gauge-free local truncation. The PEPS tensors were contracted using belief propagation with the minimal cluster size that includes the qubits being measured~\cite{Alkabetz:2020wgl,Tindall:2023dsw}.

\begin{figure}
    \centering
    \includegraphics[width=\linewidth]{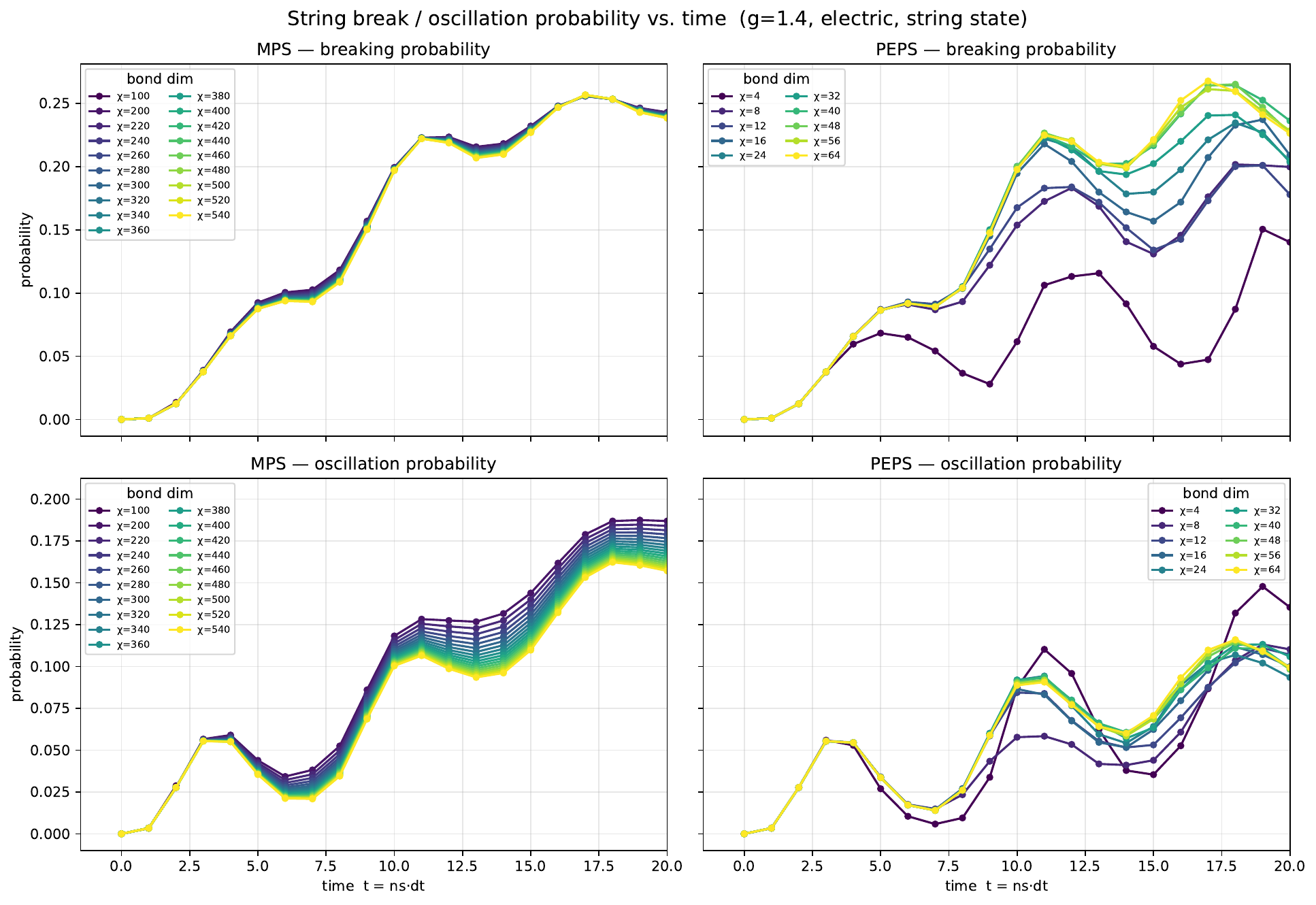}
    \caption{Breaking and oscillation probabilities for a string on an $8\times8$ lattice with $g=1.4$. The left panels show the probabilities computed using an MPS tensor network, and the right panels show the probabilities computed using a PEPS tensor network.}
    \label{fig:mps_vs_peps_prob}
\end{figure}

\begin{figure}
    \centering
    \includegraphics[width=0.5\linewidth]{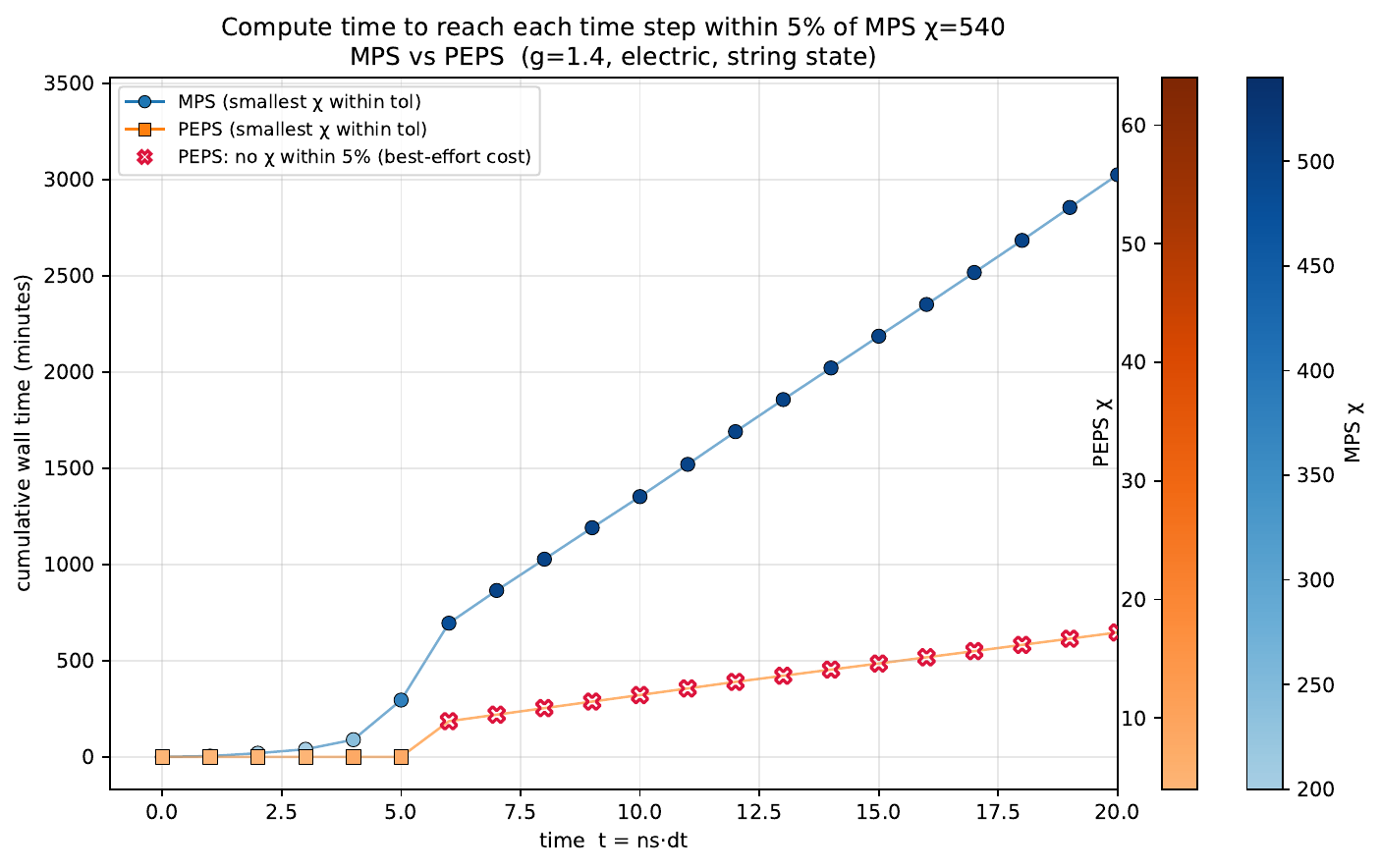}
    \caption{Walltime for the tensor network simulations performed on a single CPU Perlmutter node. For each point in time, the shown walltime corresponds to the time it took to run the minimal bond dimension that has breaking and oscillation probabilities within $5\%$ of the $\chi=540$ MPS calculation. The color on each point corresponds to the displayed bond dimension. Note that after $t=5$, no PEPS simulations are within $5\%$ of the $\chi=540$ MPS simulation.}
    \label{fig:walltimes}
\end{figure}

Fig.~\ref{fig:mps_vs_peps_prob} shows the breaking and oscillation probabilities for the string for longer times and different max bond dimensions. 
As the time evolution reaches further in time, large bond dimensions are required to maintain accuracy. Fig.~\ref{fig:walltimes} shows the wall time to reach each time step while maintaining a $5\%$ agreement in both the breaking and oscillation probabilities with the MPS simulation with a bond dimension cutoff of $540$. The MPS simulation was chosen as the reference, as it shows a clearer numerical convergence than the PEPS calculations. For the plots in the main text, the results for $\chi=100$ were used, as the MPS is fully converged at that bond dimension for the displayed time slices.
At short times, the PEPS calculations were able to give accurate results with much shorter runtimes than the MPS calculations. However, after $t=5$, the PEPS calculations are not consistent with the MPS calculations. Improving the accuracy of the PEPS calculation requires either raising the max bond dimension, utilizing a more accurate tensor network contraction strategy, or using a more accurate environment when truncating the tensor after applying a gate. Raising the bond dimension above $64$ in the PEPS calculations saturates the $476$ GB memory of a single Perlmutter node and would require a multi-node implementation of the PEPS computation. Similarly, extending the cluster radius used in the contraction beyond nearest neighbor for bond dimensions greater than $32$ exceeds the memory of a single node. The extension of the cluster to next-to-nearest neighbor for bond dimensions $\leq 32$ does not bring the results of the PEPS simulation into agreement for times past $t=5$ where agreement already exists. Note that it did shift the oscillation probabilities by an average of $2.6\%$ for times $t\geq10$ at bond dimension $\chi=32$, indicating that making precise calculations at long timescales will require going beyond the nearest-neighbor cluster. Therefore, it can be concluded that improving the accuracy of these calculations requires a better approximation of the environment of the tensors before truncating the bond dimension as well as improved contraction schemes for computing expectation values.

The $16\times8$ lattice was simulated using an MPS with max bond dimensions of $\chi=\{64,128,256,512\}$. These simulations were run on a single Perlmutter node, with each bond dimension using $32$ threads. After $8$ Trotter steps, the bond truncations of $\chi=256$ and $\chi=512$ had all link expectation values converged within $1\%$ of each other. These simulations took about $22.5$ hours to finish $8$ Trotter steps.

\bibliography{ref}

\end{document}